\documentclass[useAMS,usenatbib]{mn2e}
\usepackage{graphicx}
\usepackage{amsmath}
\usepackage{txfonts}
\usepackage{rotating}
\usepackage{lscape}
\usepackage{textcomp}

\title[Intranight polarization variability in radio-loud and radio-quiet AGN]{Intranight polarization variability in radio-loud and radio-quiet AGN}
\author[C. Villforth et al.]{Carolin Villforth$^{1,2}$\thanks{E-mail:
carovi@utu.fi}, Kari Nilsson$^{1}$, Roy \O{}stensen$^{3}$, Jochen Heidt$^{4}$, \newauthor Sami-Matias Niemi$^{1,2}$ and Janine Pforr$^{4,5}$\\
$^{1}$Tuorla Observatory, Department of Physics and Astronomy, University of Turku, V\"{a}is\"{a}l\"{a}ntie 20, 21500 Piikki\"{o}, Finland\\
$^{2}$Nordic Optical Telescope, Apartado 474, 38700 Santa Cruz de La Palma, Spain\\
$^{3}$Institute of Astronomy, K.U.Leuven, Celestijnenlaan 200D, B-3001 Heverlee, Belgium\\
$^{4}$ZAH, Landessternwarte Heidelberg, K\"{o}nigstuhl 12, 69117 Heidelberg, Germany\\
$^{5}$Institute of Cosmology and Gravitation, Mercantile House, Hampshire Terrace, University of Portsmouth, Portsmouth, PO4 0EP, United Kingdom
}
\begin{document}

\date{Accepted ?. Received ?}

\pagerange{\pageref{firstpage}--\pageref{lastpage}} \pubyear{2008}

\maketitle

\label{firstpage}

\begin{abstract}
Intranight polarization variability in AGN has not been studied extensively so far. Studying the variability in polarization makes it possibly to distinguish between different emission mechanisms. Thus  it can help answering the question if intranight variability in radio-loud and radio-quiet AGN is of the same or of fundamentally different origin. In this paper we investigate intranight polarization variability in AGN. Our sample consists of 28 AGN at low to moderate redshifts ( $0.048\leq z \leq 1.036$), 12 of which are radio-quiet quasars (RQQs) and 16 are radio-loud blazars. The subsample of blazars consists of eight flat-spectrum radio-quasars (FSRQs) and eight BL Lac objects. Each AGN was observed for a time span of $\sim4$ h in the \textit{R} band to measure polarization and variability. Using statistical methods, we determine duty cycles for polarized emission and polarization intranight variability. We find clear differences between the two samples. A majority of the radio-loud AGN show moderate to high degrees of polarization, more than half of them also show variability in polarization. There seems to be a dividing line for polarization intranight variability at $P\sim5$ per cent over which all objects vary in polarization. We did not find clear correlations between the strength of the variability and the redshift or degree of polarization. Only two out of 12 radio-quiet quasars show polarized emission, both at levels of $P<1$ per cent. The lack of polarization intranight variability in radio-quiet AGN points towards accretion instabilities being the cause for intranight flux variability whereas the high duty cycle of polarization variability in radio-loud objects is more likely caused by instabilities in the jet or changes of physical conditions in the jet plasma. We were able to constrain the timescale of the detected variations to $>4$ h. Further studies of intranight polarization variability will be necessary to reveal exact physical conditions behind this phenomenon. 
\end{abstract}


\section{INTRODUCTION}

AGN are amongst the most powerful sources of radiation in the universe. They form giant powerful jets and are thought to be key objects in the evolution of galaxies and supermassive black holes. Still, we do not yet fully understand the physics behind these objects.

The class of AGN is divided into a wide variety of subclasses. Depending on their luminosity, AGN are divided into the low-luminosity Seyfert galaxies and the high-luminosity quasars (\citealt{Veron}). AGN are also classified according to their radio-loudness into radio-loud and radio-quiet AGN, where the radio-loudness is defined as the ratio between the radio and optical flux (\citealt{Ke89}). Additionally, AGN are divided into a variety of classes according to their spectral features, polarization and variability (\citealt{UP95}).

It is an open question what is the major difference between radio-loud and radio-quiet AGN. \citet{Ul97} stated that while radio-loud AGN produce powerful collimated jets, such well collimated jets are missing in radio-quiet AGN. Differences in the spin of the the black hole might cause a AGN to be radio-loud or radio-quiet (see e.g. \citealt{Si07}). It has also been suggested that the viewing angle decides if a quasar is observed to be radio-loud or radio-quiet (\citealt{Ke89}). Yet latest studies of a radio-quiet quasar (\citealt{Bl03}) that showed a relativistically beamed jet imply that radio-loudness might be a temporal phenomenon or that the jets of radio-quiet quasars are simply far less powerful.

In this paper, we will focus on two classes of AGN: radio-quiet quasars (RQQs) and blazars. The blazar class is divided into two subclasses: BL Lac objects and flat-spectrum radio quasars (FSRQs). While in BL Lacs the emission lines are either weak or absent, FSRQs show broad emission lines. Both types are characterised by a flat radio spectrum (\citealt{Ul97}). So far, all known blazars are radio-loud (\citealt{AS80}). According to the standard interpretation, blazars are AGN with a jet pointing directly towards the observer (\citealt{UP95}). Radiation from the jet is thought to be highly beamed due to the small viewing angle. One of the most striking properties of this class is their violent variability on all timescales from hours to decades, see e.g. \citet{EVa00}, \citet{St05}. Also, blazars are known to show extremely high degrees of polarization and variability. In many cases, the most variable objects also show the highest degrees of polarization (\citealt{AS80}). This makes them promising objects for studying intranight polarization variability. A detailed review of polarization properties of blazars can be found in \citet{AS80}. Based on gamma-ray data, \citet{Gh09} suggested that there might be a clear difference between FSRQs and BL Lacs, with FSRQs showing more efficient accretion and thus stronger jets, while the accretion power is weak in BL Lacs, resulting in less collimated jets and weak line emission.

In this paper, we will study the variability in polarization on time scales of hours, called intranight polarization variability. Studying short time scale flux variations in AGN, two different terms are used to specify the presence or absence of variability: microvariability and intranight variability. While microvariability refers to variability on $mmag$ levels, usually on smaller time scales (\citealt{MN96}), intranight variability refers to variability on time scales of hours (\citealt{WW95}). Thus, microvariability is defined by the strength of the variations while intranight variability is defined by the time scale of the variations. In this study, we will use the concept of intranight variability to define the presence or absence of variability.

Intranight variability have been observed in numerous AGN (see e.g. \citealt{WW95}, \citealt{GoK00}). The phenomenon is most prominent in blazars and radio-loud quasars (see e.g.\citealt{WW95}). However, even though they show less violent variability, also radio-quiet quasars (RQQs) have been found to intranight variability (see e.g. \citealt{St05}, \citealt{GoK00}). Still, the origin of intranight variability is not yet established.

If one wants to understand the origin of intranight variability in AGN, one needs to distinguish between the different mechanisms that can cause the variability. A wide variety of theoretical models has been proposed to explain variations on small time scales. A way to distinguish between these models is using statistical arguments, as different mechanisms are fostered by certain conditions. 

Instabilities in the jet are thought to cause violent variability in radio-loud AGN, especially in the subclass of blazars (\citealt{WW95}). Different models exist that explain variability in jet emission, e.g. through lighthouse effects (see e.g. \citealt{GKW92}), shock fronts in jets (see e.g. \citealt{MG85}, \citealt{Ma08}) or certain physical conditions in the jet plasma (see e.g. \citealt{CIB85}).

The ultraviolet wavelength region in quasars is dominated by thermal emission from the accretion disk, the so-called \textquoteleft blue bump\textquoteright (\citealt{Ga08}). Thus, variability in quasars at these wavelengths is more likely to be caused by accretion instabilities. This dependency can be studied by observing quasars at different redshifts with the same filter and thus at different restframe wavelength. However, this makes it possible that any actual redshift dependency in the variability properties of quasars might be interpreted as wavelength dependency.

For example, it has been suggested by \citet{La08} that in the optical bands, RLQs might have a higher contribution from the non-thermal component than RQQs. They studied a large sample of RQQs and RLQs and were able to model the differences in the spectral energy distributions (SED) assuming different spectral indexes for the non-thermal component, such that RQQs are dominated by thermal emission in the optical, while for the RLQs the non-thermal component is dominant. Thus, comparing the duty cycle and strength of INOV in RLQs and RQQs can help answering the question if INOV is intrinsic to the non-thermal or thermal component, i.e. the jet or disk emission.

A phenomenon which has not been studied elaborately so far in this context is polarization INOV (PINOV, referring to variability in polarization on time-scales of hours). \citet{An05} studied intranight polarization variability in BL Lacs, they found very high duty cycles for both polarization and polarization variability. Polarized emission in AGN can have different reasons. An important origin of polarized emission are relativistic jets. Due to the strong magnetic fields in relativistic jets, the emission from the jet can be dominated by synchrotron emission and is thus highly polarized (see e.g. \citealt{BB82}). Due to unclear mechanisms, not only the degree of polarization, but also the position angle can change on relatively small time scales (see e.g. \citealt{Ho84}). It is not yet clear what causes the changes in position angle (see e.g. \citealt{CIB85}). Also in nearby AGN, polarization due to scattering in the extended gas region around the nucleus or the disk has been detected (\citealt{An82}). Different models predict more or less low polarized emission from scattering in the accretion disk (see e.g. \citealt{La90}, \citealt{AB96}, \citealt{GG07}). However, polarization due to scattering is not expected to cause a rotating position angle. Additionally, if the distribution of the position angle of polarization is symmetric it will cancel out when summing up over the whole object. Thus, this kind of polarized emission is unlikely to be observed at high redshifts.

While the observation of flux variability does not give a direct hint towards the mechanisms that cause it, polarization variability does. Thus, by studying the polarization variability and their link to flux variability in different types of AGN, it is possible to find out which mechanisms causes the variations. This might be the only direct way to examine if intranight variability in radio-loud and radio-quiet AGN is caused by the same physical conditions or if the origin of intranight variability in radio-loud and radio-quiet AGN is fundamentally different.

Our project combines the usage of polarization measurements and samples of radio-loud and radio-quiet AGN at different redshift. We measure the duty cycle of PINOV and it's strength and derive statistical differences between the two samples.

In Section 2 we describe the sample, followed by a description of the observations in Section 3. The data reduction is presented in Section 4, Section 5 sums up the results of our study, followed by Discussion in Section 6 and finally Conclusions in Section 7.

\section{SAMPLE}

We selected the AGN according to the following criteria from the V\'{e}ron-Cetty \& V\'{e}ron catalogue (\citealt{Veron}): AGN were selected to have a listed apparent magnitude in the \textit{R} band of $m_{R}<16$ mag to ensure high enough signal-to-noise with exposure times in the range $<60$ s at the 2.5-m Nordic Optical Telescope (NOT). In order to minimise dilution of possible variability or polarization due to host galaxy flux, we require a non-detection of the host galaxy or $M_{host} > M_{nucleus} + 2$. For objects where such data was not available, we assumed a host galaxy absolute magnitude $M^{*}$ and applied the same criterion. Additionally, we required at least one comparison star at similar magnitude within the FOV of the ALFOSC calcite plate (circle of $\sim140$\textacutedbl diameter).

\begin{figure}
\includegraphics[width=8cm]{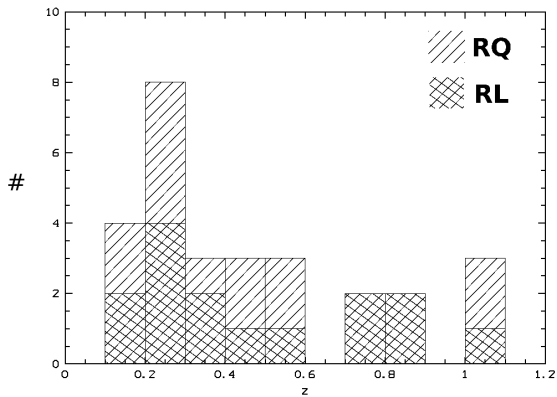}
\caption{Redshift distribution of the quasar sample. RQQs are shown as striped surfaces, radio-loud objects (FSRQ and BL Lac type objects) as checkered surfaces.}
\label{zdistribution}
\end{figure}

Our sample consists of 28 AGN, 16 of them are radio-loud, 12 are radio-quiet. Among the radio-loud AGN, eight are BL Lac type objects and eight are FSRQs (Flat Spectrum Radio Quasars). The redshift range for the sample is $0.048\leq z \leq 1.036$ ($0.048\leq z \leq 1.024$ for the radio-quiet objects and $0.138\leq z \leq 1.036$ for the radio-loud objects). Table \ref{sample} shows the sample. The redshift distribution of our sample, separately for the radio-loud and radio-quiet AGN is shown in Fig. \ref{zdistribution}.

\begin{table*}
\begin{minipage}{180mm}
\caption{The Quasar-Sample. Date of Observations refers to beginning of night. References for identification: a) \citet{Veron}, b) \citet{PG95}, c) \citet{JaFSRQ}}
\label{sample}
\centering
\begin{tabular}{l c c c c c c c}
\hline \hline
ID & Date of Observations & $z$ & RA & DEC & $m_{R}$ & type  \\
\hline
PG 0026+129     & 27.09.2006 & 0.145 & 00:29:13.69 & +13:16:03.9 & $\sim15$ & RQQ $^{a}$\\
PG 0043+039     & 26.09.2006 & 0.384 & 00:45:47.27 & +04:10:24.4 & $\sim16$ & RQQ $^{a}$\\
FBQS J0242+0057 & 25.09.2006 & 0.569 & 02:42:40.32 & +00:57:27.2 & $\sim16$ & RQQ $^{a}$\\
1E 0514-0030    & 07.02.2007 & 0.291 & 05:16:33.50 & -00:27:13.5 & $\sim16$ & RQQ $^{a}$\\
1ES 0647+250    & 21.01.2006 & 0.203 & 06:50:46.60 & +25:03:00.0 & $\sim16$ & BL Lac $^{a,b}$\\
S5 0716+714     & 20.01.2006 & 0.300 & 07:21:53.45 & +71:20:36.4 & $\sim15$ & BL Lac $^{a,b}$\\
PKS 0735+178    & 08.02.2007 & 0.424 & 07:38:07.39 & +17:42:19.0 & $\sim15$ & BL Lac $^{a,b}$\\
QSO B0754+100   & 06.02.2007 & 0.266 & 07:57:06.64 & +09:56:34.9 & $\sim15$ & BL Lac $^{a,b}$\\
1ES 0806+524    & 13.03.2008 & 0.138 & 08:09:49.19 & +52:18:58.4 & $\sim15$ & BL Lac $^{a,b}$\\
1WGAJ0827.6+0942& 14.03.2008 & 0.260 & 08:27:40.10 & +09:42:10.0 & $\sim15$ & RQQ $^{a}$\\
OJ 287          & 21.01.2006 & 0.306 & 08:54:48.87 & +20:06:30.6 & $\sim14$ & BL Lac $^{a,b}$\\
QSO B0953+414   & 23.04.2007 & 0.239 & 09:56:52.41 & +41:15:22.1 & $\sim15$ & RQQ $^{a}$\\
3C 232          & 16.03.2008 & 0.530 & 09:58:20.95 & +32:24:02.3 & $\sim16$ & FSRQ $^{a,c}$\\
FBS 0959+685    & 17.03.2008 & 0.773 & 10:03:06.77 & +68:13:16.9 & $\sim15$ & FSRQ $^{a,c}$\\
OM 280          & 06.02.2007 & 0.200 & 11:50:19.21 & +24:17:53.8 & $\sim16$ & BL Lac $^{a,b}$\\
OM 295          & 15.03.2008 & 0.729 & 11:59:31.83 & +29:14:43.8 & $\sim15$ & FSRQ $^{a}$\\
ON 325          & 13.03.2008 & 0.237 & 12:17:52.08 & +30:07:00.6 & $\sim14$ & FSRQ $^{a,c}$\\
3C 273          & 08.02.2007 & 0.158 & 12:29:06.70 & +02:03:08.6 & $\sim12$ & FSRQ $^{a,c}$\\
PG 1246+586     & 18.03.2008 & 0.847 & 12:48:18.78 & +58:20:28.7 & $\sim16$ & BL Lac $^{a,b}$\\
3C 279          & 24.04.2007 & 0.536 & 12:56:11.17 & -05:47:21.5 & $\sim12$ & FSRQ $^{a}$\\
PG 1254+047     & 25.04.2007 & 1.024 & 12:56:59.93 & +04:27:34.4 & $\sim16$ & RQQ $^{a}$\\ 
PG 1259+593     & 14.03.2008 & 0.472 & 13:01:12.93 & +59:02:06.7 & $\sim15$ & RQQ $^{a}$\\
PG 1307+085     & 07.02.2007 & 0.155 & 13:09:47.03 & +08:19:49.3 & $\sim16$ & RQQ $^{a}$\\
CSO 873         & 18.03.2008 & 1.014 & 13:19:56.23 & +27:28:08.2 & $\sim15$ & RQQ $^{a}$\\
PKS 1656+053    & 25.04.2007 & 0.879 & 16:58:33.45 & +05:15:16.4 & $\sim15$ & FSRQ $^{a,c}$\\
PG 1700+518     & 24.04.2007 & 0.292 & 17:01:24.87 & +51:49:21.0 & $\sim14$ & RQQ $^{a}$\\
PG 2112+059     & 25.09.2006 & 0.457 & 21:14:52.57 & +06:07:42.5 & $\sim16$ & RQQ $^{a}$\\
CTA 102         & 27.09.2006 & 1.036 & 22:32:36.41 & +11:43:50.9 & $\sim16$ & FSRQ $^{a,c}$\\
\hline
\end{tabular}
\end{minipage}
\end{table*}

\section{OBSERVATIONS}

Observations were carried out at the Nordic Optical Telescope on La Palma (Canary Islands, Spain) using ALFOSC, the Andalucia Faint Object Spectrograph and Camera, and FAPOL. FAPOL is an integral part of the FASU unit used in combination with ALFOSC. The observations were done in visitor mode on 5 separate runs in 2006 -- 2008. For the observations we used ALFOSC in its polarimetric mode. The instrument setup consists of a $\lambda/2$ wave plate inside FAPOL and a Calcite in the aperture wheel of ALFOSC that splits the incoming light into an ordinary and an extraordinary beam. The ALFOSC filter wheel that we used is located after the Calcite in the lightpath, thus the filter does not introduce errors to the polarization measurements. We used four positions of the $\lambda/2$ wave plate: $0^{\circ}, 22.5^{\circ}, 45.0^{\circ}$ and $67.5^{\circ}$, corresponding to a rotation of the plane of the electric field of $0^{\circ}, 45^{\circ}, 90^{\circ}$ and $135^{\circ}$. We used the fast photometry mode of ALFOSC which makes it possible to quickly read out several windows on the chip (see \citealt{OS00}). Therefore, we only read out the two beams for the object and all available comparison stars and one additional sky window for each object. As the Calcite used in ALFOSC is relatively small, it produces a field-of-view of about $\sim140$\textacutedbl  in diameter, thus the number of available comparison stars is low. Depending on the object, we were able to observe one to three comparison stars. As the single windows are of the size $~13$\textacutedbl$\times13$\textacutedbl, we did not use standard sky subtraction for the photometry using a sky annulus. Instead, we used the sky window that was set directly next to the two beams of the corresponding objects for sky subtraction. Exposure times were chosen depending on the magnitude of the object and the comparison stars, such that the read out noise did not dominate the errors for a single measurement. We used the \textit{R} band for all observations. The instrumental polarization was checked using zero-polarization standards and is negligible. Additionally, possible instrumental polarization will be subtracted using the field stars. However, this does not correct for possible variations of the instrumental polarization over the field-of-view. For other instruments it was found that the instrumental polarization is caused by asymmetry in the light path (see e.g. \citealt{PR06}). As the field-of-view when using the Calcite is extremely small, the level of asymmetry in the light path is low and therefore, the instrumental polarization should be stable within the field-of-view. 

\section{DATA REDUCTION}

Data were reduced using ESO-MIDAS \footnote{European Southern Observatory Munich Image Data Analysis System} and rtcnv\footnote{See http://www.not.iac.es/instruments/alfosc/fastphot/rtcnv-guide.html}. Data in fast photometry mode are saved as a series of FITS files containing a one dimensional pixel stream and a map of which physical CCD pixels are read out. The rtcnv program is used to produce two-dimensional FITS images that can be analysed with regular image processing software. We also used rtcnv to average the overscan pixels in order to determine the bias level, which was subtracted from the image. This corrects time variation of the bias level. The ALFOSC CCD does not have any spatial structure in the bias level, thus subtracting a bias level and not a bias frame is fully sufficient.

In principle flatfielding is not necessary when using dual beam polarimeters as the flatfielding effects cancel out in case the object does not move on the chip within a cycle of four exposures that form one polarimetric measurement. However, we tested different flatfielding methods to check for possible improvements. We discuss different flatfielding methods for polarimetry and their advantages and disadvantages in the following paragraphs.

We did not use flatfields with the Calcite and retarder due to the following reasons: imbalances in ALFOSC cause bending of the instrument when moving the telescope. This causes apparent movement of the dust specks on the chip. One way to use flatfields with Calcite and retarder in the beam without introducing errors would be to stop the observations about every $30$ min to take dome flats and continue observations afterwards. However, this would require closing and reopening the dome, which takes about $10$ min. Additionally, during that time guiding would not be possible. Blind tracking might cause the star to move out of the box, causing additional overheads for re-acquisition of the target. Since any gain would be lost by the large overheads, this is clearly not a useful option. Considering the fact that polarimetry flatfields are not as simple as photometry flatfields as they show two overlapping fields, it should additionally be kept in mind that these corrections are non-trivial.

On the other hand, it is unclear how well flatfields with only the filter in the beam work. It is unclear how well they correct for large-scale variations as the lightpath changes significantly when inserting the calcite and retarder. We performed tests concerning this matter and found that this kind of flatfields correct at least for the very large scale variations, i.e. illumination patterns. However, they do definitely not correct for dust specks due to the strong differences in the lightpath. On the other hand, pixel to pixel variations, i.e. differences in the gain between the pixels, should be corrected by these kind of flatfields. We performed tests to check this assumption. We checked the scatter of the measured normalised Stokes parameters for flatfielded and non-flatfielded data. Results where inconclusive, some targets showed slight increase, some slight decrease in scatter when using flatfielded data. It seems that either the gain differences between the pixels or the exact positioning of the windows on the chip are not completely stable. It is also possible that the gain differences are sensitive to the polarization of the detected light. Thus, we decided not to flatfield the data.

As we do not flatfield our data, we expect problems with the photometry. Even we use guiding for our observations, the objects move on the chip during the 4 hours of the observations, usually on scales of $~1$--$5$ pixels. This changes the influences of dust specks and thus can cause false variability, especially if we consider that microvariability might occur on scales of $mmag$. We performed tests on the two objects that have three comparison stars (1E 0514-0030 and CTA 102). In both cases variability is detected in all magnitude measurements. Thus, we conclude that photometry from polarimetric data is not reliable if proper flatfielding cannot be performed. This can be a serious problem for polarimetry as well. In some cases, not the degree of polarization but the polarized flux might be of interest a. If reliable photometry is not available, errors in polarized flux occur.

The flux was measured using aperture photometry. The flux in each beam was measured using circular apertures, the sky was subtracted using the corresponding sky region. We skipped the first and last ten pixels of the sky regions due to read-out problems at the beginning of each window showing as an increasing bias level. We tested two methods for the sky determination: the median value and the mode value determined as follows:

\begin{equation}
mode = 3 \times median - 2 \times mean
\end{equation}

The mode value should avoid wrong sky determination in presence of faint objects in the sky field. These might not be visible when inspecting the single frames, but might show in a skewed distribution of sky values. We ran the data reduction with the two different sky value determination methods and checked for differences in the scatter of the normalised Stokes parameters. The mode sky determination method yielded better results. We measured a clear decrease in the scatter for more than 95 per cent of the normalised Stokes parameters. We thus use the mode sky determination for all our data.

The best aperture was determined by running the reduction for a wide range of apertures, we then choose the aperture that yields the lowest scatter in the normalised Stokes parameters $P_{x}$ and $P_{y}$ for the main target. In case of bad data, i.e. clouds passing, almost fully diluting the object, we reject these bad data by setting a limit for the signal-to-noise-ratio. This method yields moderately to significantly bigger apertures than optimisation using the signal-to-noise-ratio in the target. As our study is clearly focused on polarimetry and not photometry we decided to optimise all data reduction for this purpose. 

The degree of polarization and position angle of polarization were calculated as follows:
\begin{equation}
Q_{i} = \dfrac{E_{i}}{O_{i}}
\end{equation}
\begin{equation}
Q_{M} = Q_{0^{\circ}} + Q_{22.5^{\circ}} + Q_{45^{\circ}} + Q_{67.5^{\circ}}
\end{equation}
\begin{equation}
P_{x} = \dfrac{Q_{0^{\circ}} - Q_{45^{\circ}}}{Q_{M}}
\end{equation}
\begin{equation}
P_{y} = \dfrac{Q_{22.5^{\circ}} - Q_{67.5^{\circ}}}{Q_{M}}
\end{equation}
\begin{equation}
P = \left( \sqrt{P_{x}^{2} + P_{y}^{2}} \right) \times 100\%
\end{equation}
\begin{equation}
PA = 28.7 \times atan2\left(P_{y},P_{x} \right)
\end{equation}
With $E_{i}$ and $O_{i}$ being the flux in the extraordinary and ordinary beam for the angles $i=0^{\circ}, 22.5^{\circ}, 45^{\circ}, 67.5^{\circ}$, respectively. For the error calculations, we included shot and read-out noise. Errors in polarization were calculated using error propagation for the above formulae.

We subtracted background polarization using the field stars. This assumes that both the quasars and the corresponding field stars are affected by the same background polarization. This assumption was tested by comparing the polarization of different field stars within one field, if available. In most of the cases all field stars within one field yielded the same degree of polarization and position angle within error limits. However, this estimate of background polarization yields only a lower limit. Additional background polarization caused by dust aligned with the Galactic magnetic field might be present. This might for example be the case if the comparison stars are nearby objects and are affected by less absorption and thus less background polarization. We will comment on this in the result section for the objects for which the determination of the background polarization was critical. Background polarization was subtraced vectorially, i.e. in $P_{x}$ and $P_{y}$ with $P_{x/y}$ being the normalised Stokes parameters.

Data averaging for the polarization and position angle was performed by averaging $P_{x}$ and $P_{y}$ and calculating the averaged polarization from the background subtracted values. Errors for averaged polarization were calculated from the scatter of $P_{x}$ and $P_{y}$ using error propagation. Additionally, we considered the bias due to non-gaussian distribution of the degree of polarization from \citet{SS85}. The unbiased degree of polarization was calculated as follows:

\begin{equation}
P_{unbiased} = \sqrt{P_{bg subtracted}^{2} - 1.41^{2} \times \sigma_{P_{bg subtracted}}^{2}} 
\end{equation}
We consider all objects for which the unbiased degree of polarization $P_{unbiased}$ is consistent with zero polarization to be unpolarized. For degrees of polarization $P < 2$ per cent we give an estimate of the asymmetric behaviour of the error distributions derived from \citet{SS85}. 

We apply zero polarization angle correction using measurements of high polarization standards. This correction was only applied to the final averaged data to allow comparison with other data. The calibrated position angle of polarization $PA$ is listed in Table \ref{resulttab}.

Variability detection is challenging. Normal $\chi^{2}$-tests compare the scatter with the mean calculated error. This poses the problem that faulty error calculations cause faulty detections of variability. For the photometry we use error estimates that are standard for aperture photometry as used in IRAF\footnote{Image Reduction and Analysis Facility, distributed by NOAO, operated by the AURA, Inc. under agreement with the NSF}, however other sources of error than shot  and read-out noise are expected. For polarimetry the case is even more severe. \citet{PR06} studied elaborately the influences of systematic errors in polarimetric measurements. These systematic errors are found to depend on the instrument, the telescope and the number of retarder plate positions used. Other causes of systematic errors may exist. Additionally, the error propagation used only considers shot and read-out  noise. Incomplete error estimates for the photometry thus also cause underestimation of the errors for polarimetry. As the detection of variability is the main goal of this study, normal $\chi^{2}$-tests as described above are not sufficient.

We use the analysis of variance (ANOVA) to detect PINOV. This statistical test is robust and does not rely on error estimations but uses the scatter of the data to detect variability. Additionally, ANOVA has been used by other authors (see e.g. \citealt{deD98}, \citealt{St05}) that studied INOV in RLQs and RQQs.

We perform a one way ANOVA using the stats package from SciPy\footnote{see http://www.scipy.org}. We divide the data into bins of ten data points. In our result table we give the probability that the data in the different bins are not drawn from the same parent population, i.e. the probability $p_{var}$ that the object is variable. For $p_{var} > 95$ per cent in either $P_{x}$ or $P_{y}$ the object is considered as variable. We perform ANOVA on the values of $P_{x}$ and $P_{y}$ as the degree of polarization shows non-gaussian distribution at low degrees of polarization. ANOVA explicitly assumes Gaussian distributions, thus it cannot be used for the degree of polarization.

Additionally, we inspect all data by eye. Some of the data where ANOVA clearly favours variability show weird variability patterns, i.e. after averaging and plotting the data, we see a single heavy outlier and otherwise stable normalised Stokes parameters. For such cases we perform the following reliability check. We remove the datapoints that cause the single heavy outlier and rerun the ANOVA on the rest of the data. If the second ANOVA does not detect variability, the variability is \textquoteleft dubious\textquoteright.

We present the results of the statistical tests and other important properties, i.e. the average polarization, the number of comparison stars and the exposure time in Table \ref{resulttab}. The plots of the polarization variability are shown in the Fig. \ref{firstplot} to \ref{lastplot}.

\section{RESULTS}

\begin{figure}
\includegraphics[width=8cm]{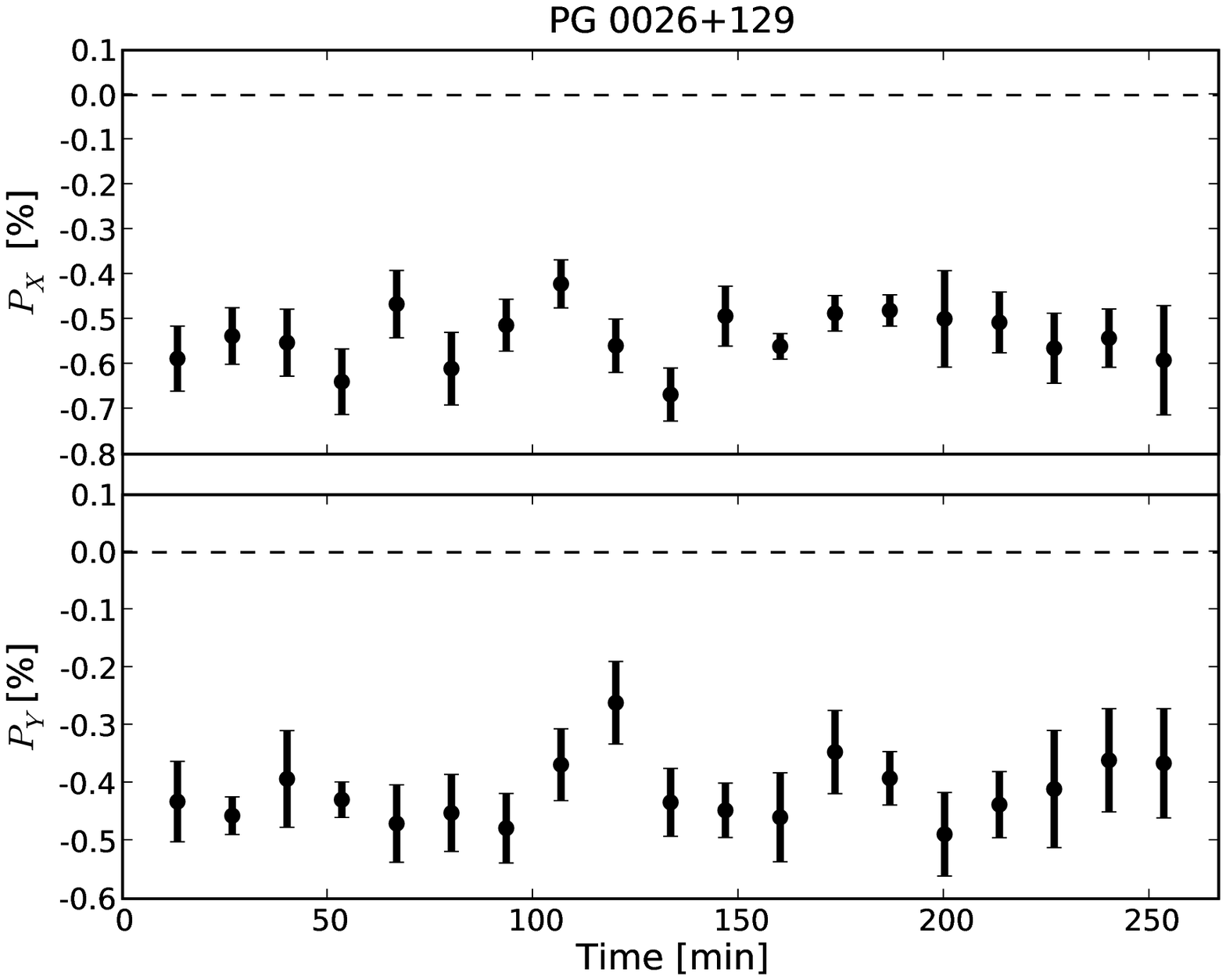}
\vspace{0.5cm}
\includegraphics[width=8cm]{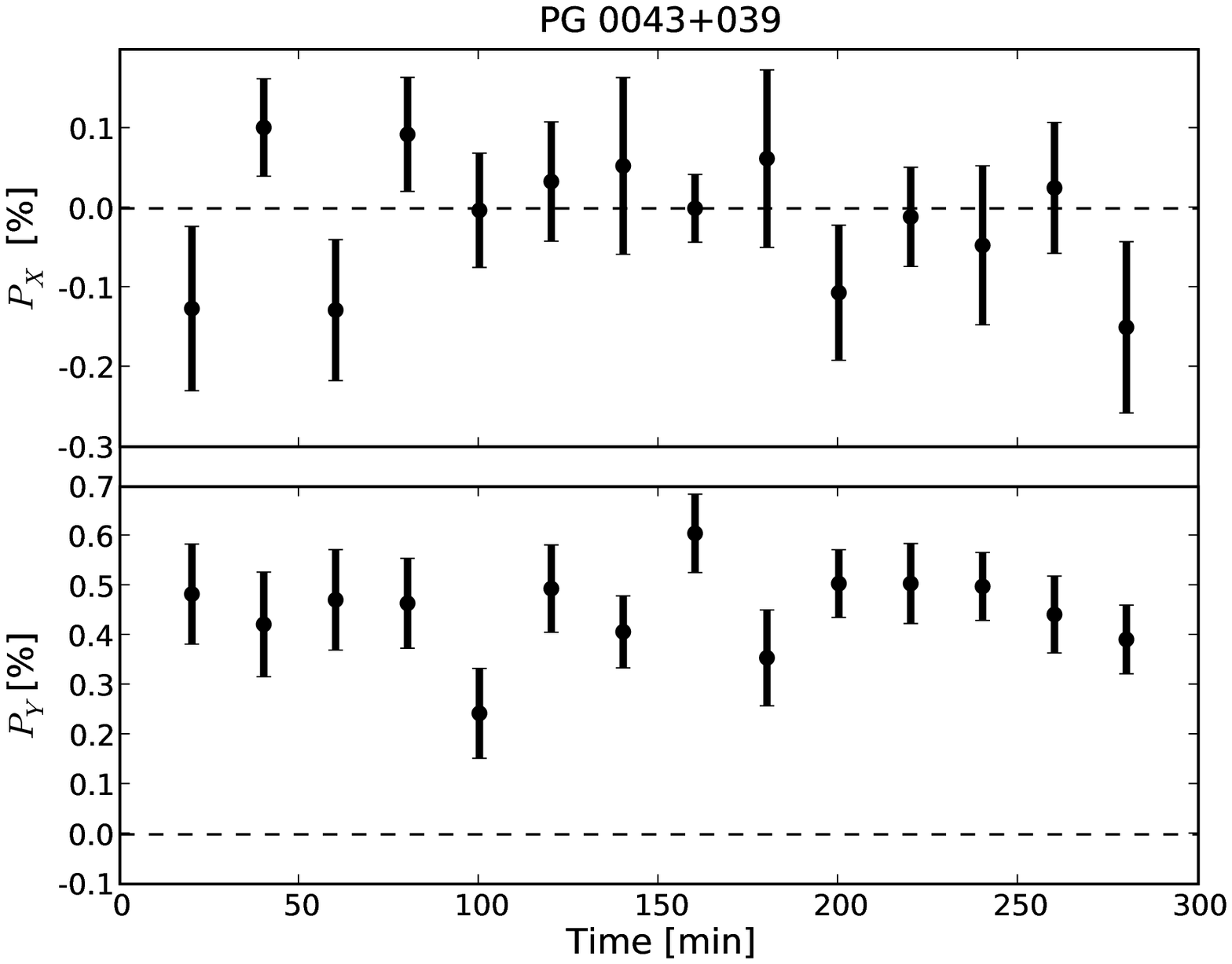}
\vspace{0.5cm}
\includegraphics[width=8cm]{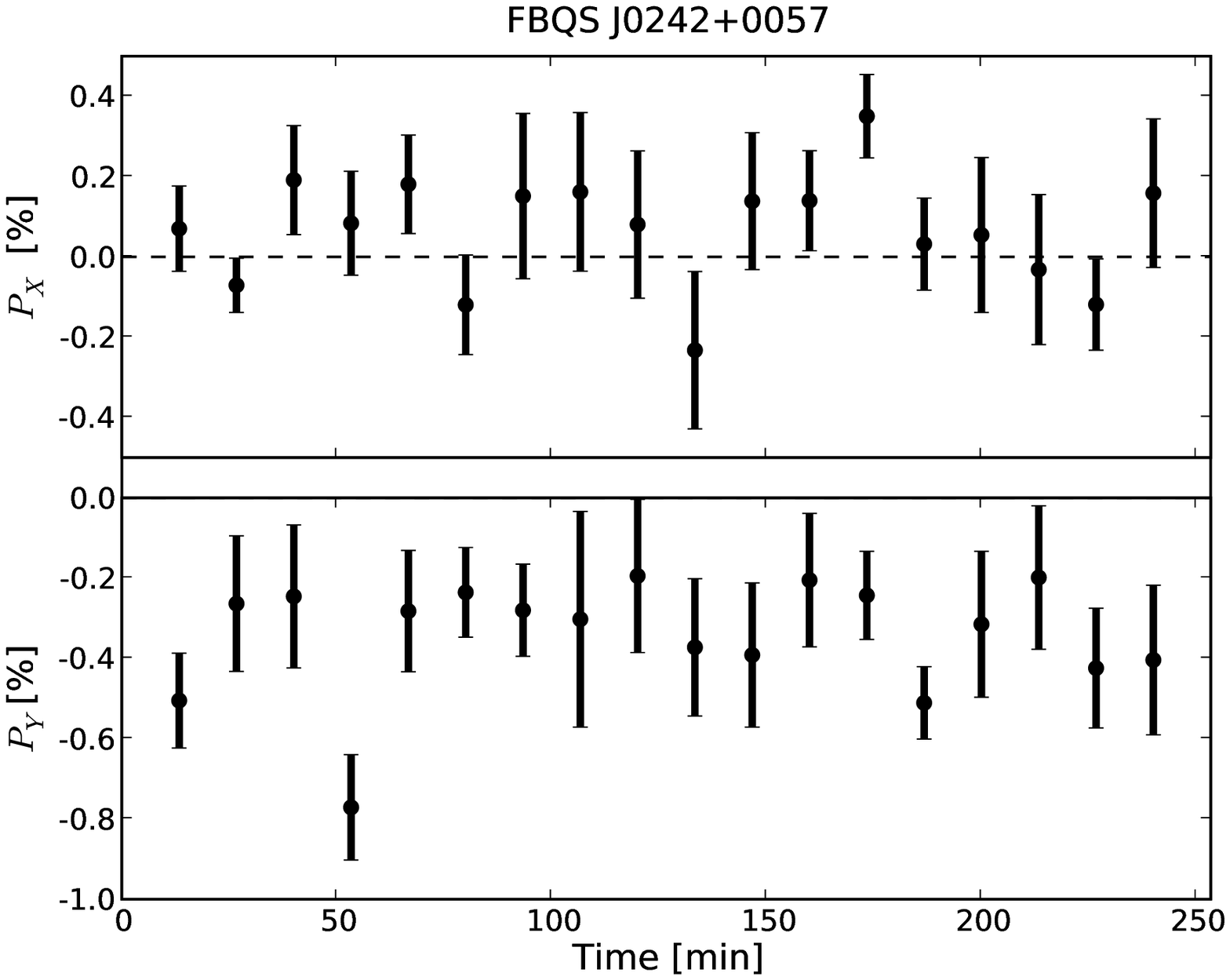}
\caption{Polarization intranight variability. We applied a signal to noise limit for each object. The resulting data were binned with a bin size of ten. Data were averaged. We plot the mean and standard error. Dashed horizontal lined indicate $P_{x/y} = 0$. We plotted normalised Stokes parameters $P_{x}/P_{y}$ against time in minutes. Background polarization is not subtracted. Plots continue in Fig.\ref{nextplot} to \ref{lastplot}.}
\label{firstplot}
\end{figure}

\begin{figure}
\includegraphics[width=8cm]{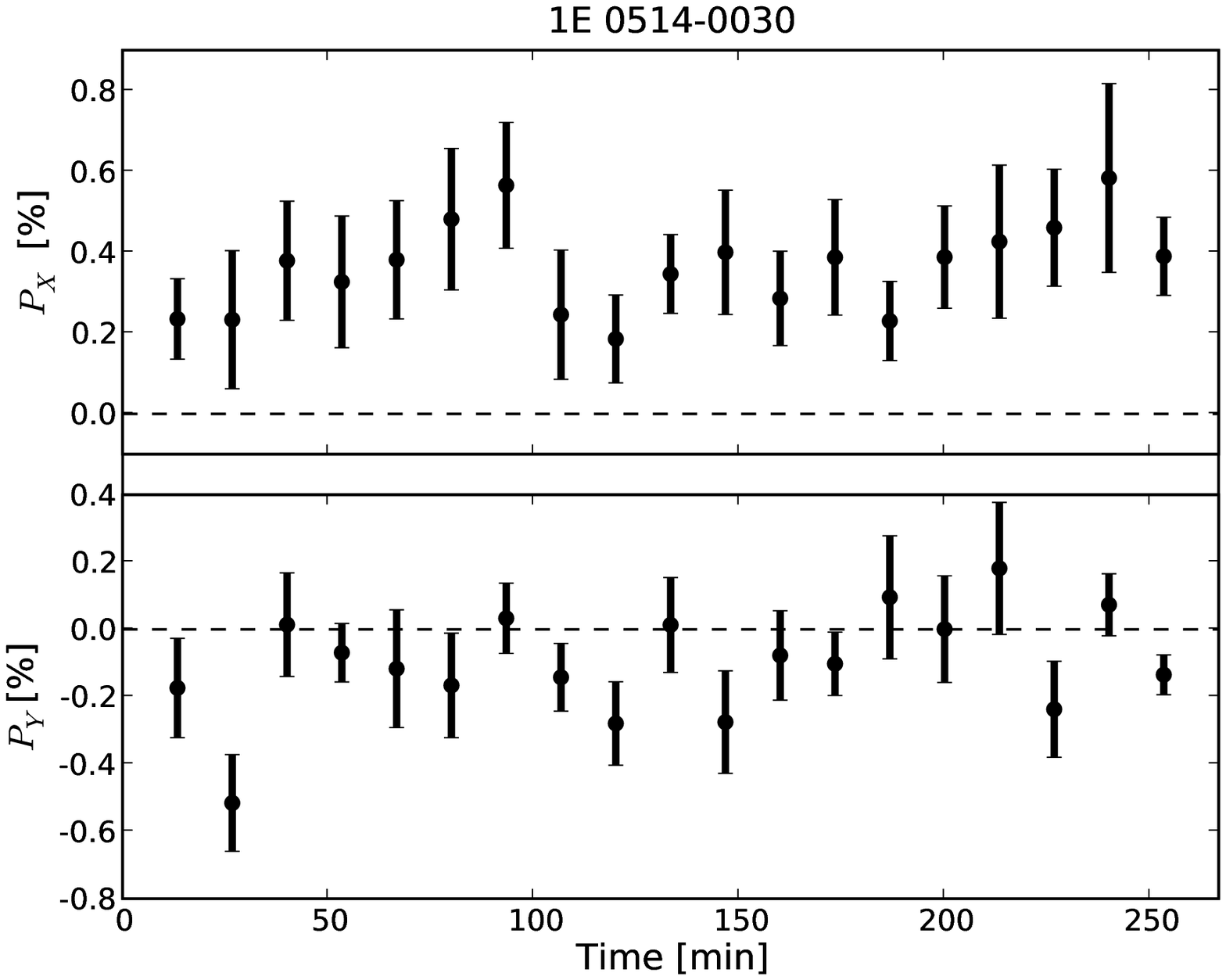}
\vspace{0.5cm}
\includegraphics[width=8cm]{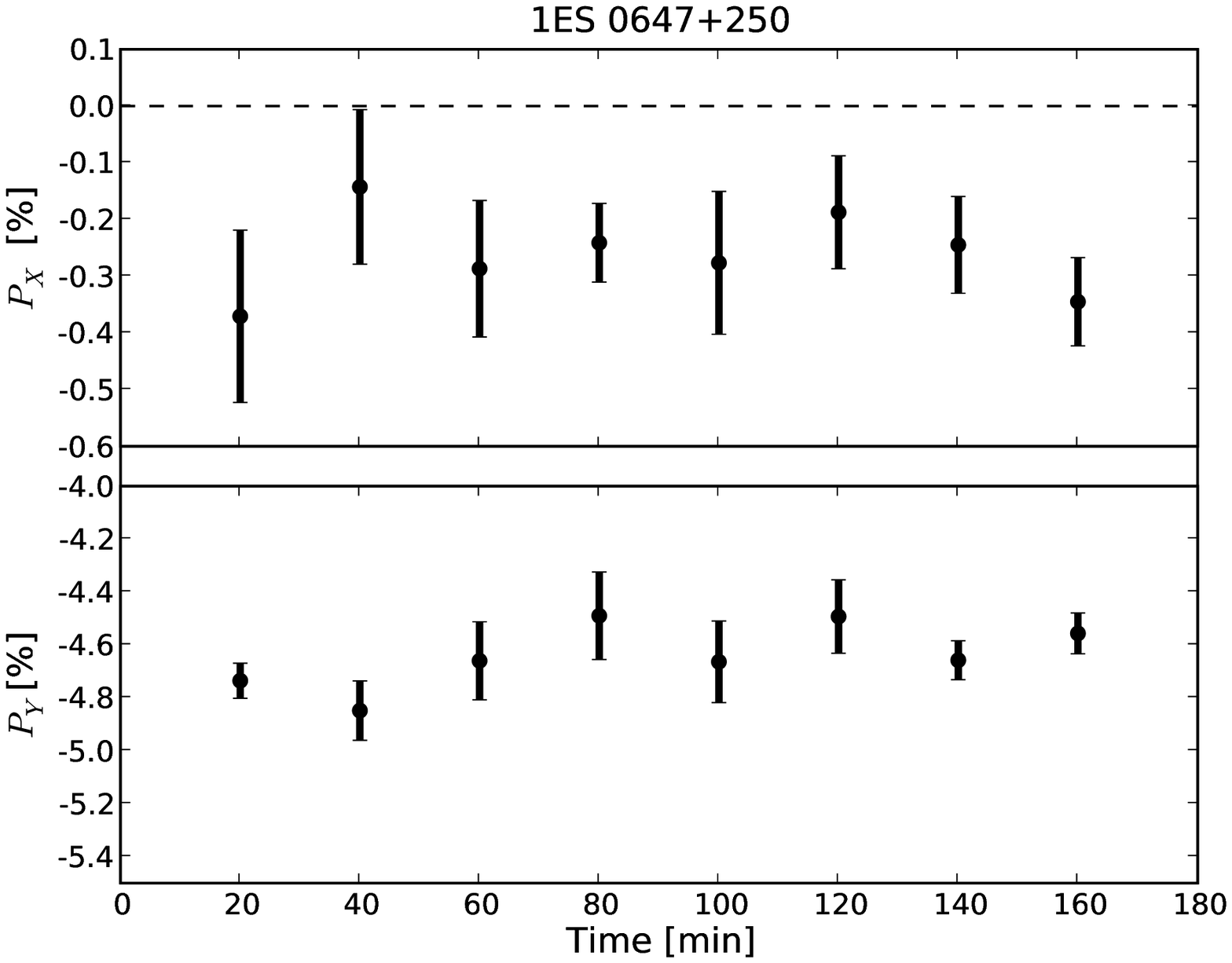}
\vspace{0.5cm}
\includegraphics[width=8cm]{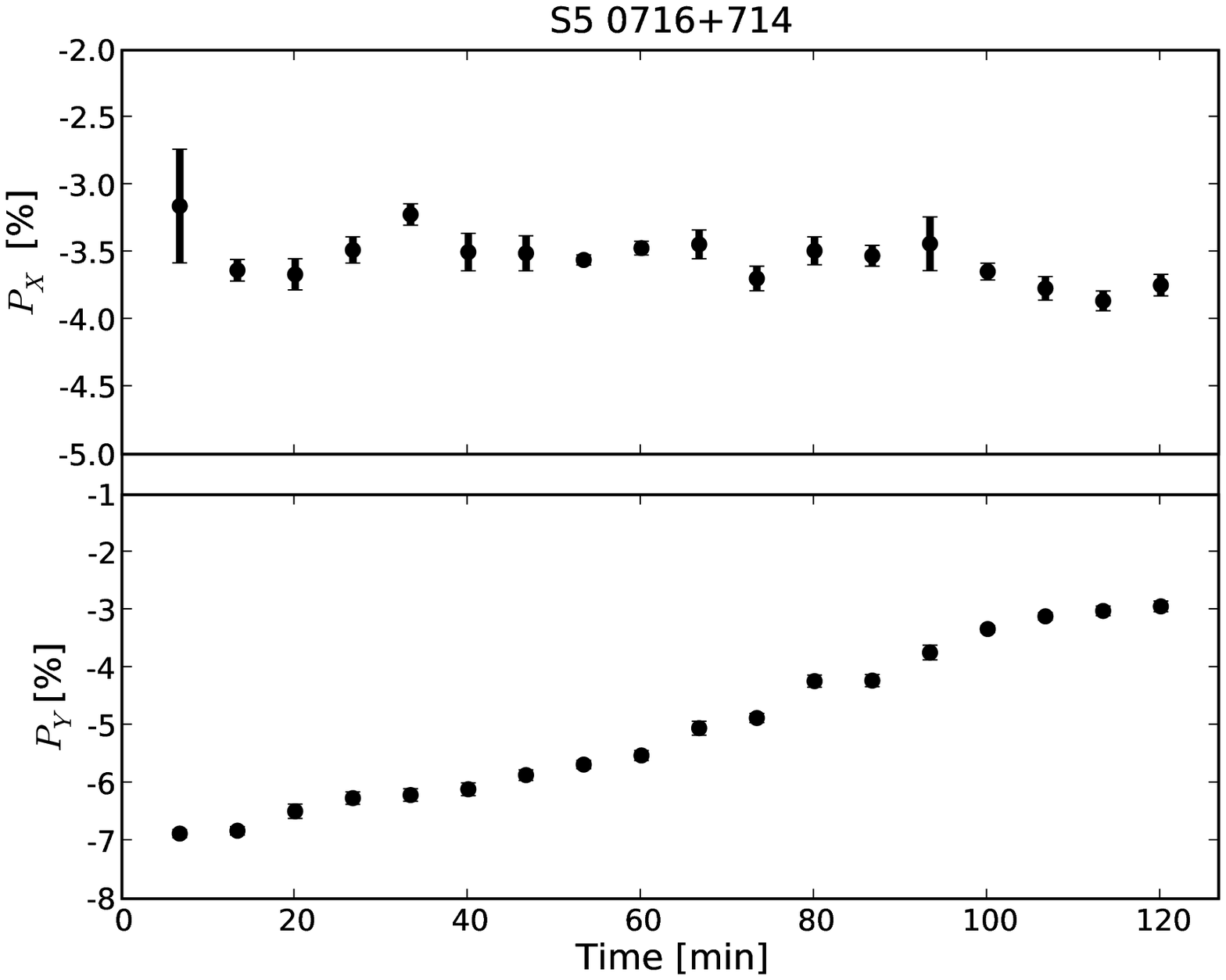}
\caption{Polarization intranight variability, continued. For caption see Fig. \ref{firstplot}.}
\label{nextplot}
\end{figure}

\begin{figure}
\includegraphics[width=8cm]{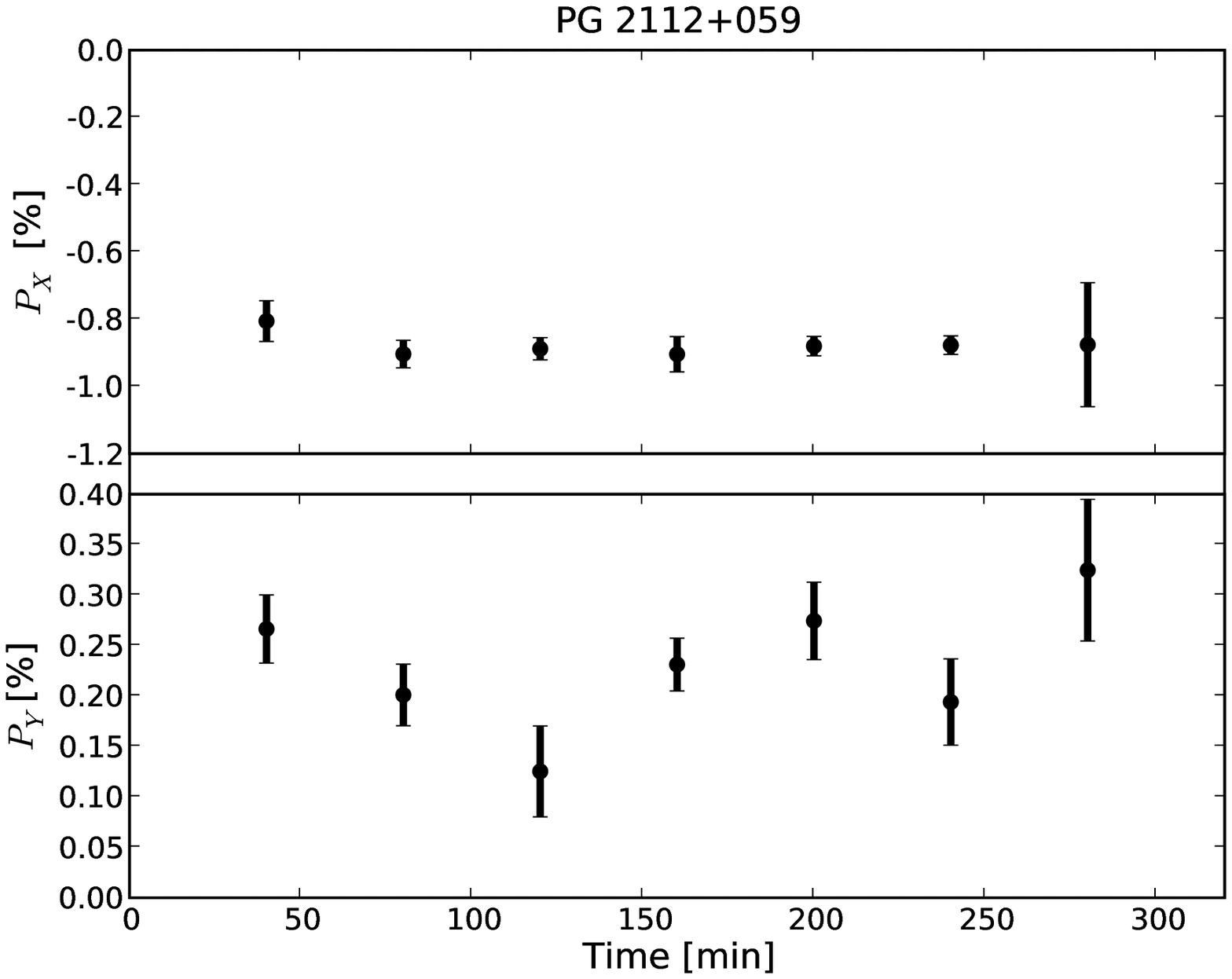}
\vspace{0.5cm}
\includegraphics[width=8cm]{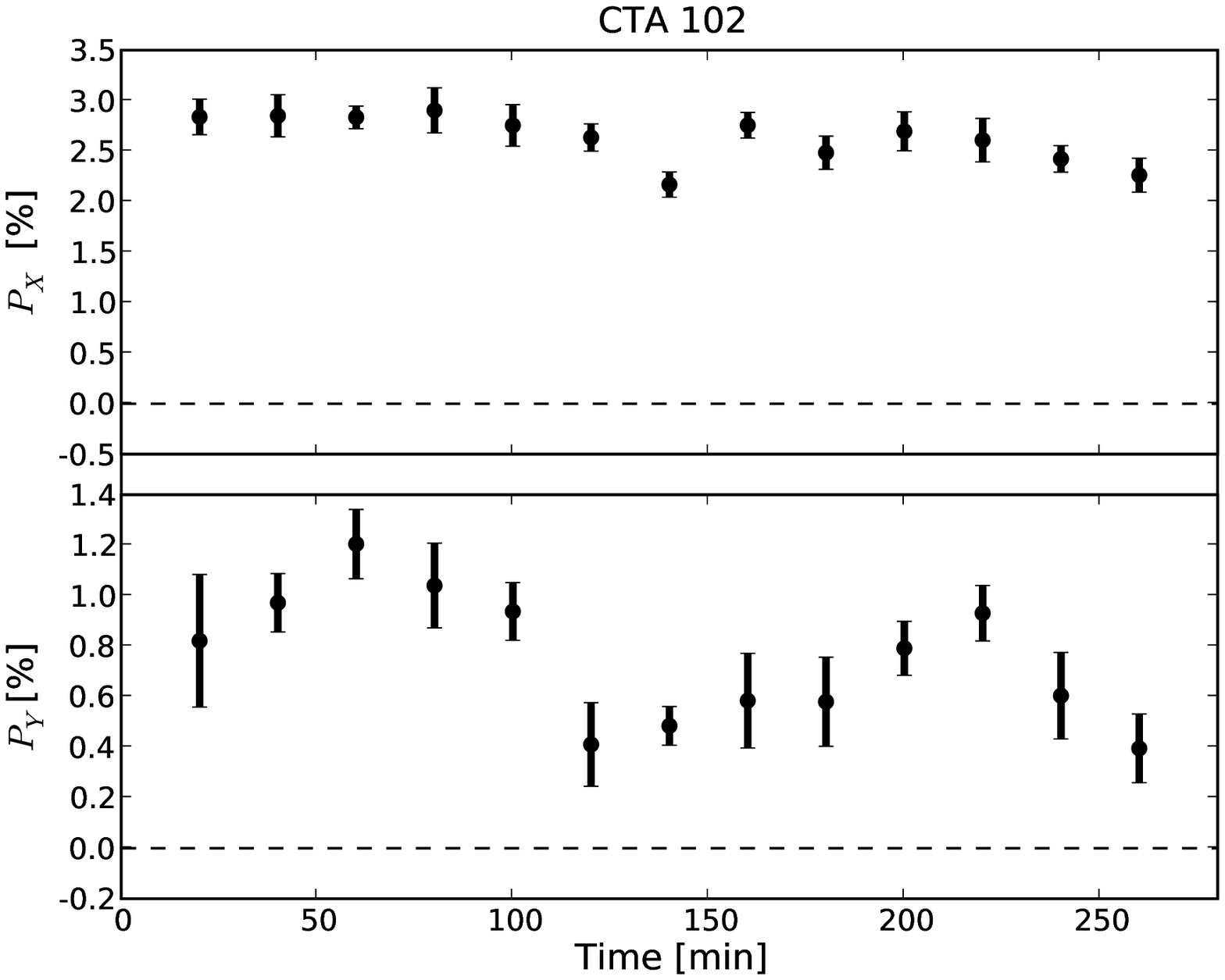}
\caption{Polarization intranight variability, continued. For caption see Fig. \ref{firstplot}.}
\label{lastplot}
\end{figure}

In the following paragraphs we discuss the results for selected targets. We do not present the results for targets that are unpolarized and do not show variable polarization. Results for all targets can be found in Table \ref{resulttab}. Plots of normalised Stokes parameters $P_{x}$ and $P_{y}$ for all targets are presented in Figures \ref{firstplot} to \ref{lastplot}.

\begin{figure}
\includegraphics[width=8cm]{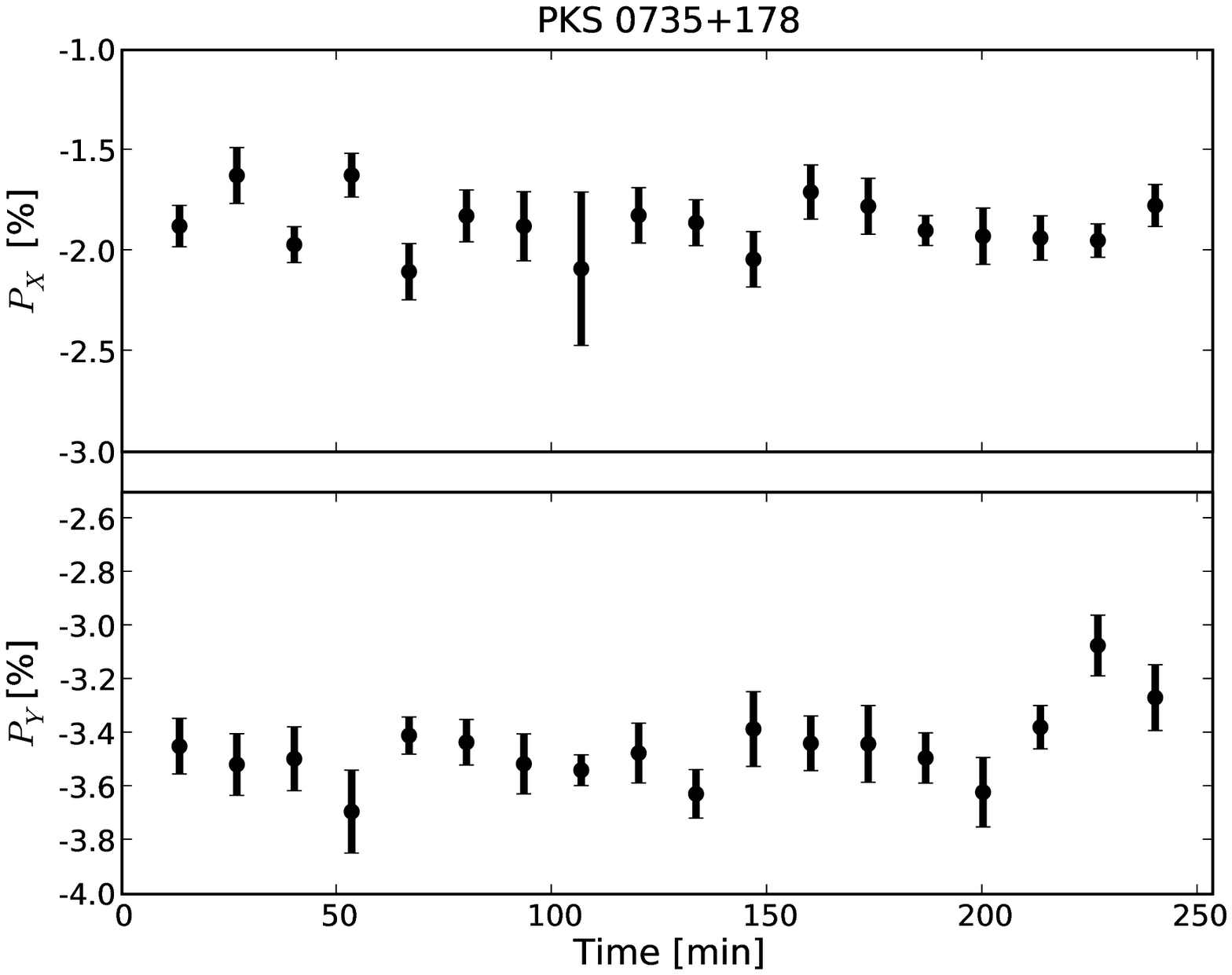}
\vspace{0.5cm}
\includegraphics[width=8cm]{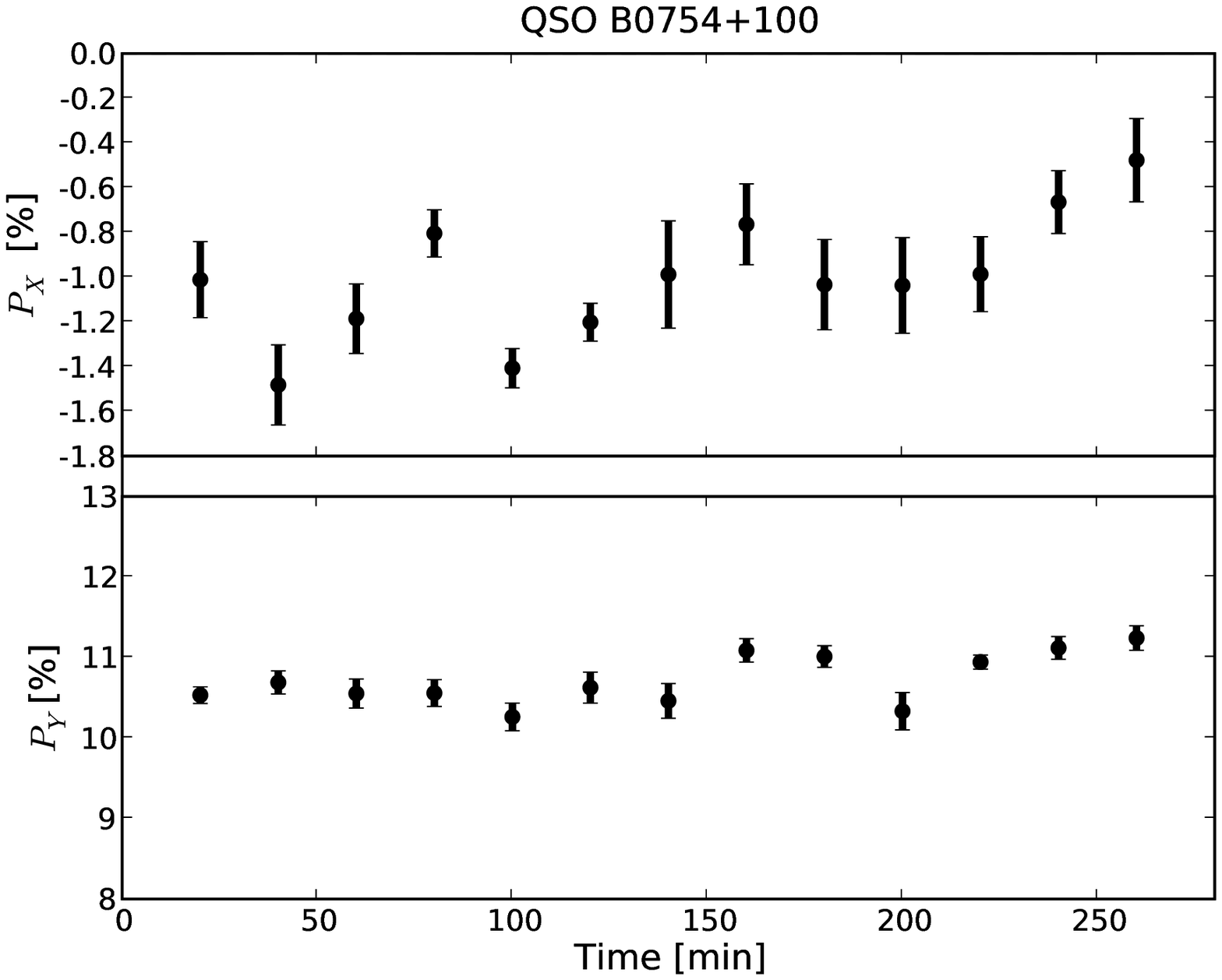}
\vspace{0.5cm}
\includegraphics[width=8cm]{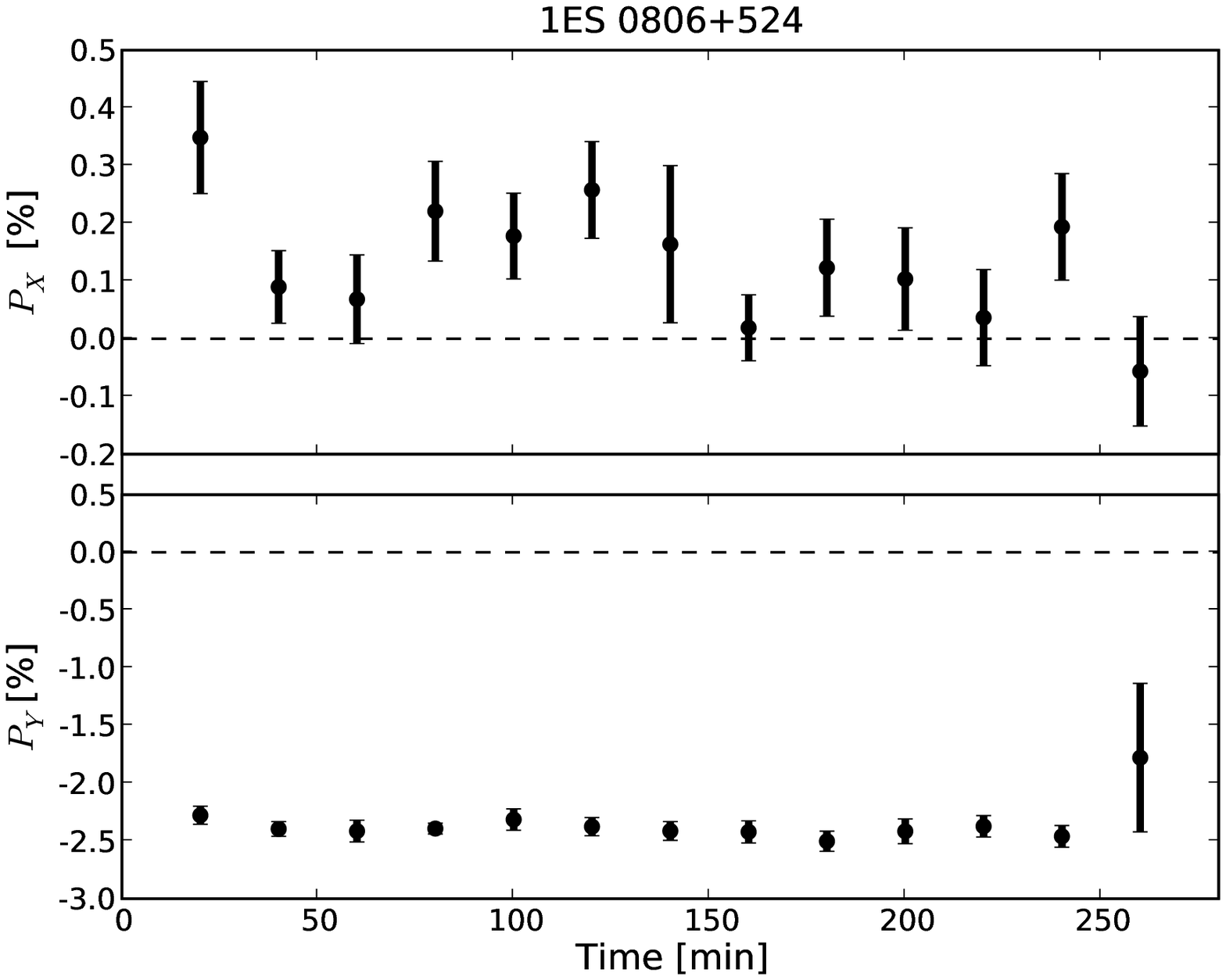}
\caption{Polarization intranight variability, continued. For caption see Fig. \ref{firstplot}.}
\end{figure}

\begin{figure}
\includegraphics[width=8cm]{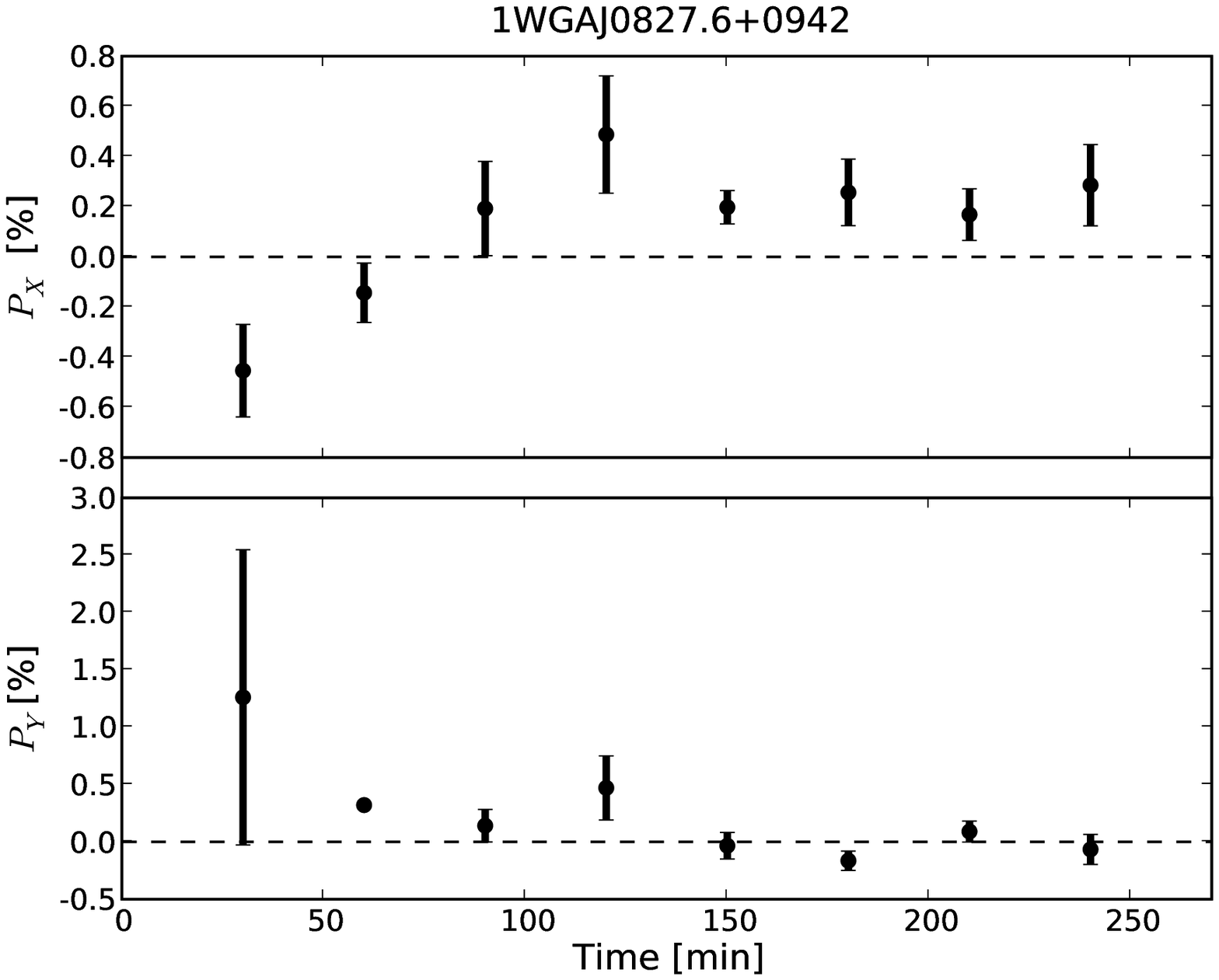}
\vspace{0.5cm}
\includegraphics[width=8cm]{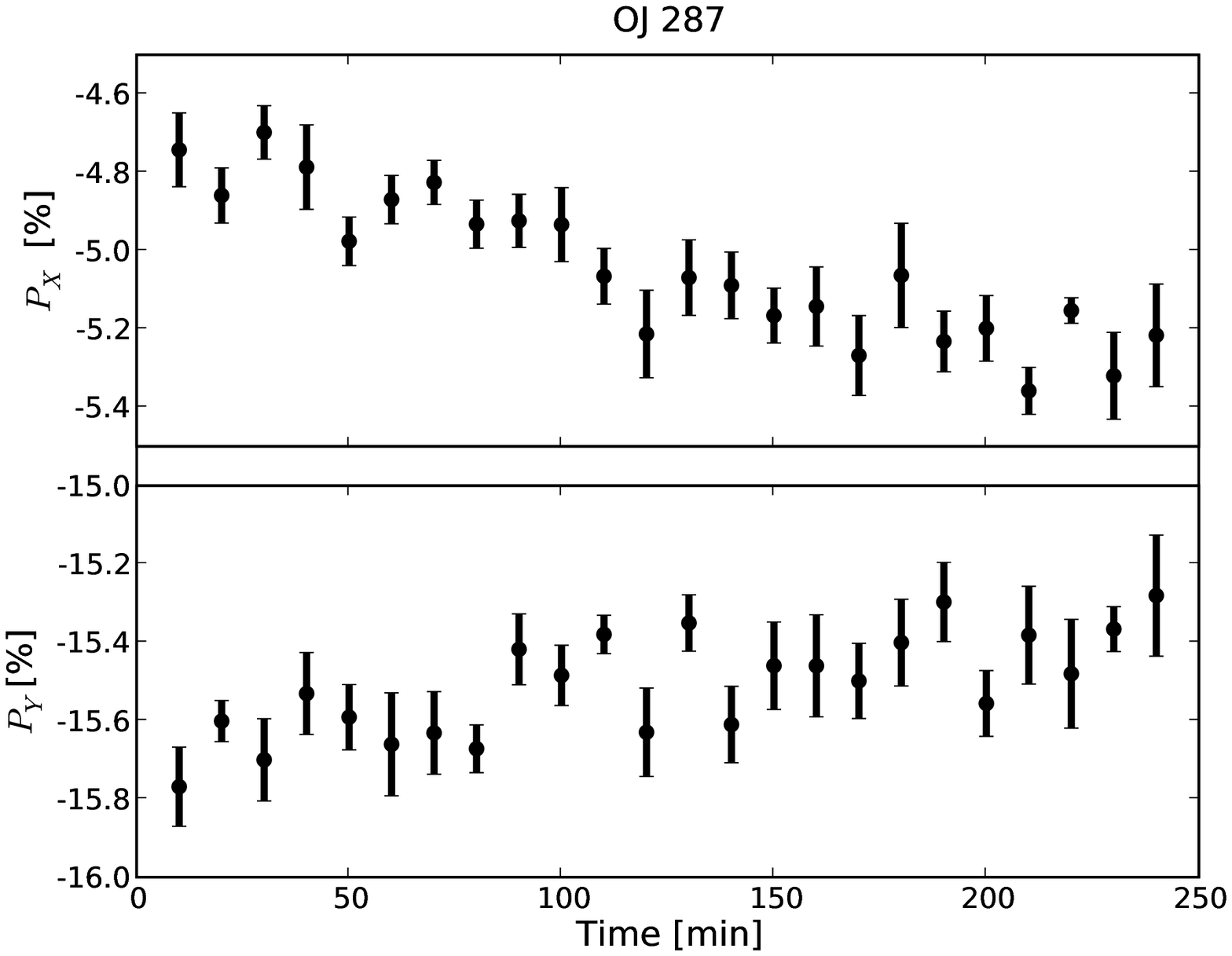}
\vspace{0.5cm}
\includegraphics[width=8cm]{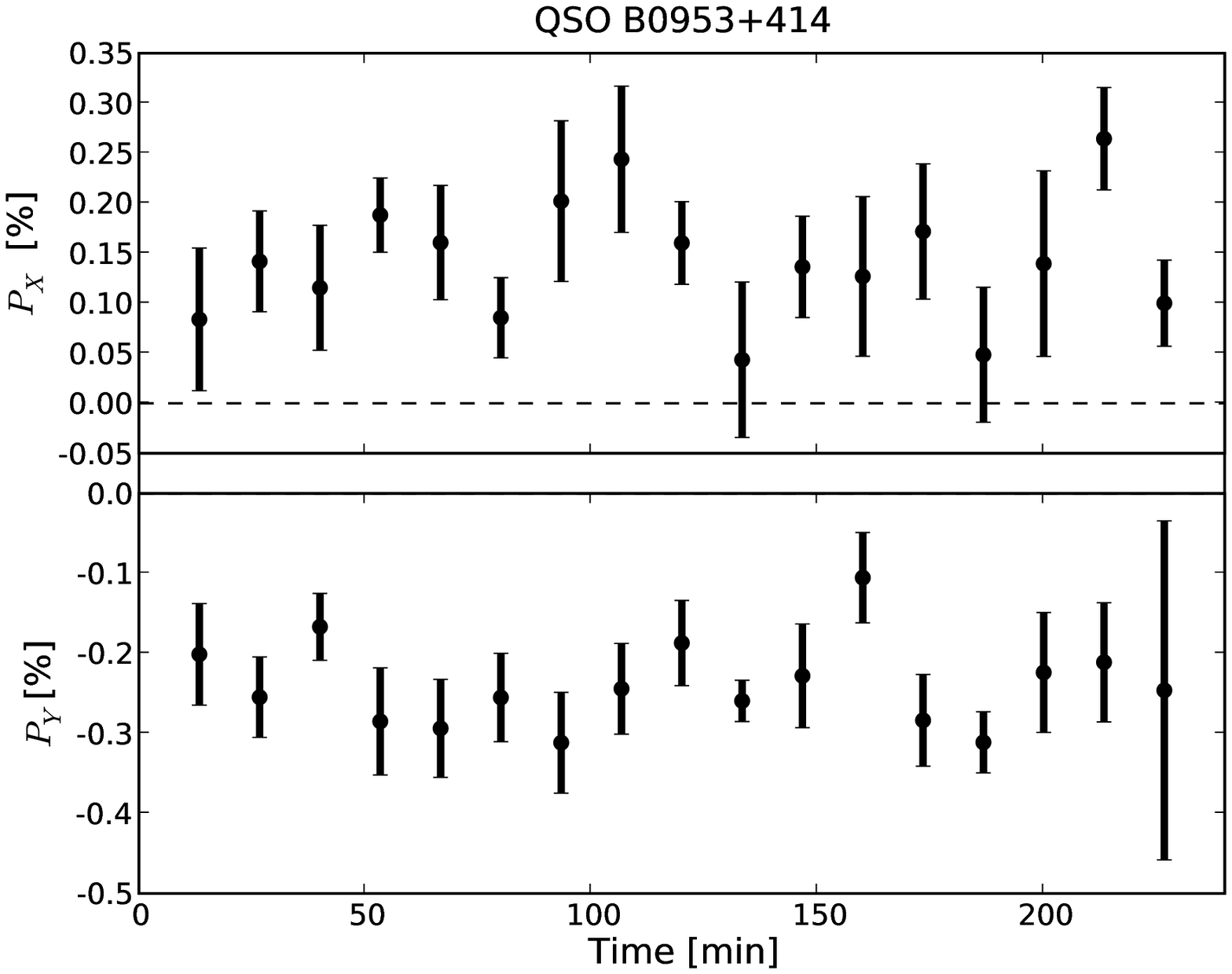}
\caption{Polarization intranight variability, continued. For caption see Fig. \ref{firstplot}.}
\end{figure}

\begin{figure}
\includegraphics[width=8cm]{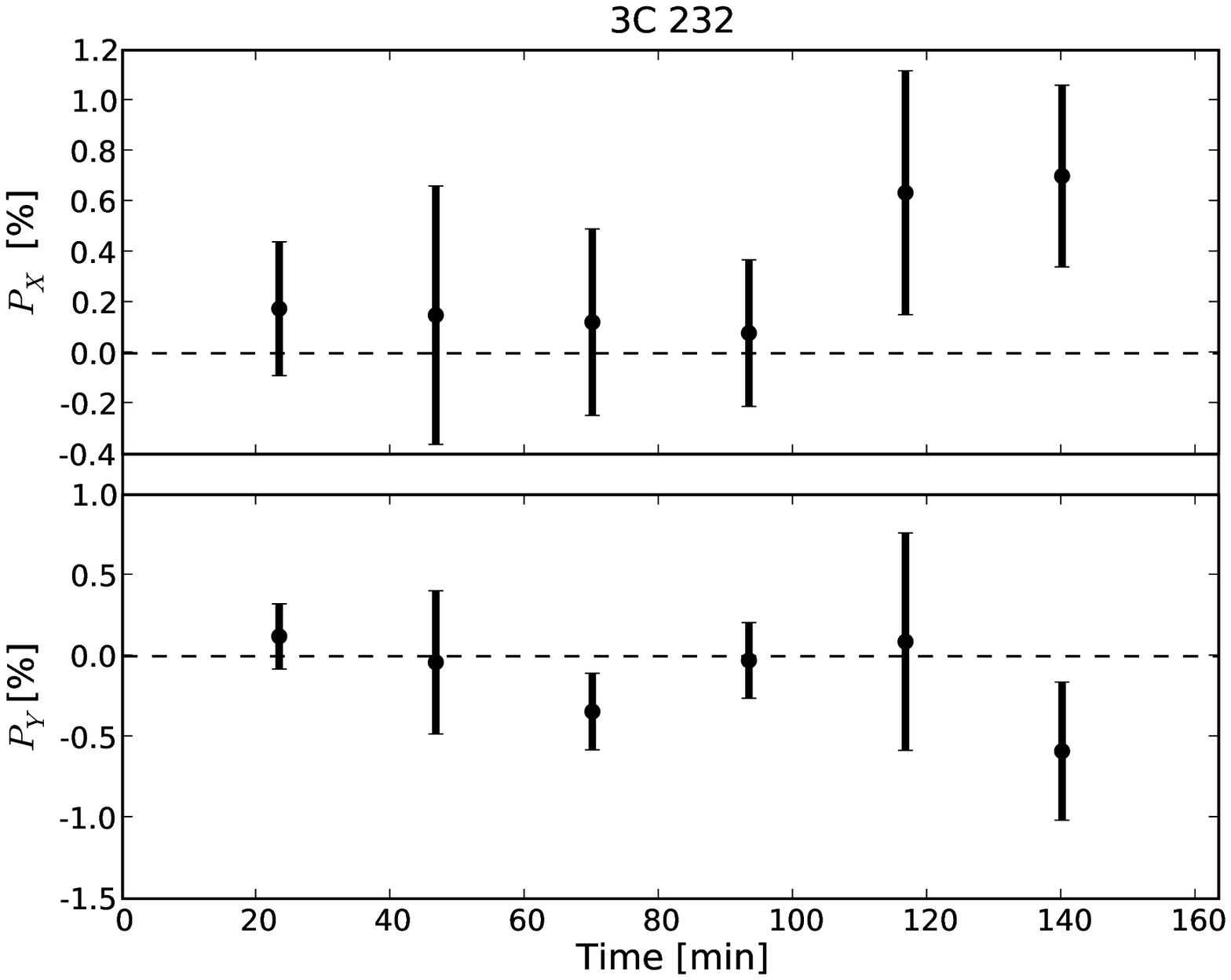}
\vspace{0.5cm}
\includegraphics[width=8cm]{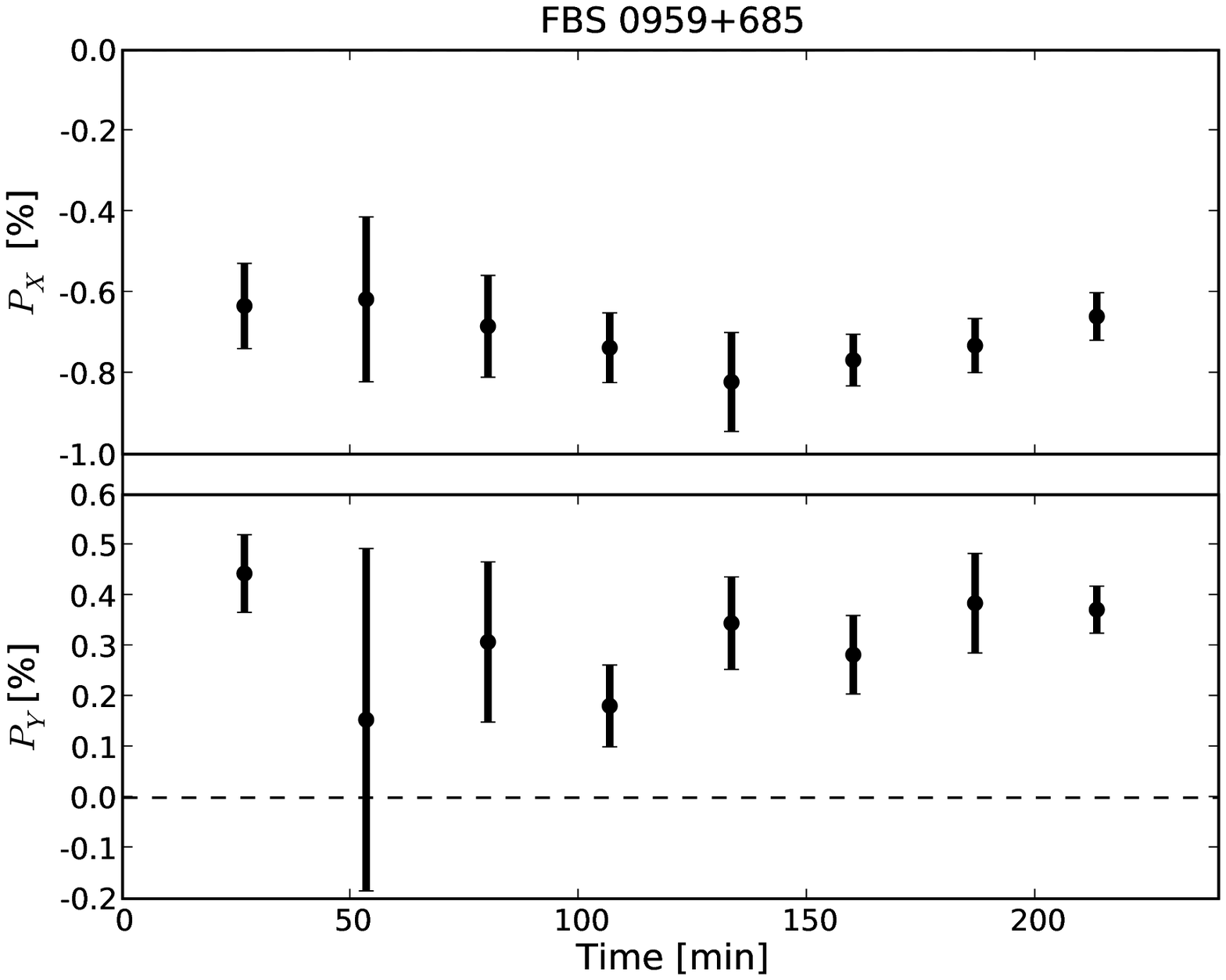}
\vspace{0.5cm}
\includegraphics[width=8cm]{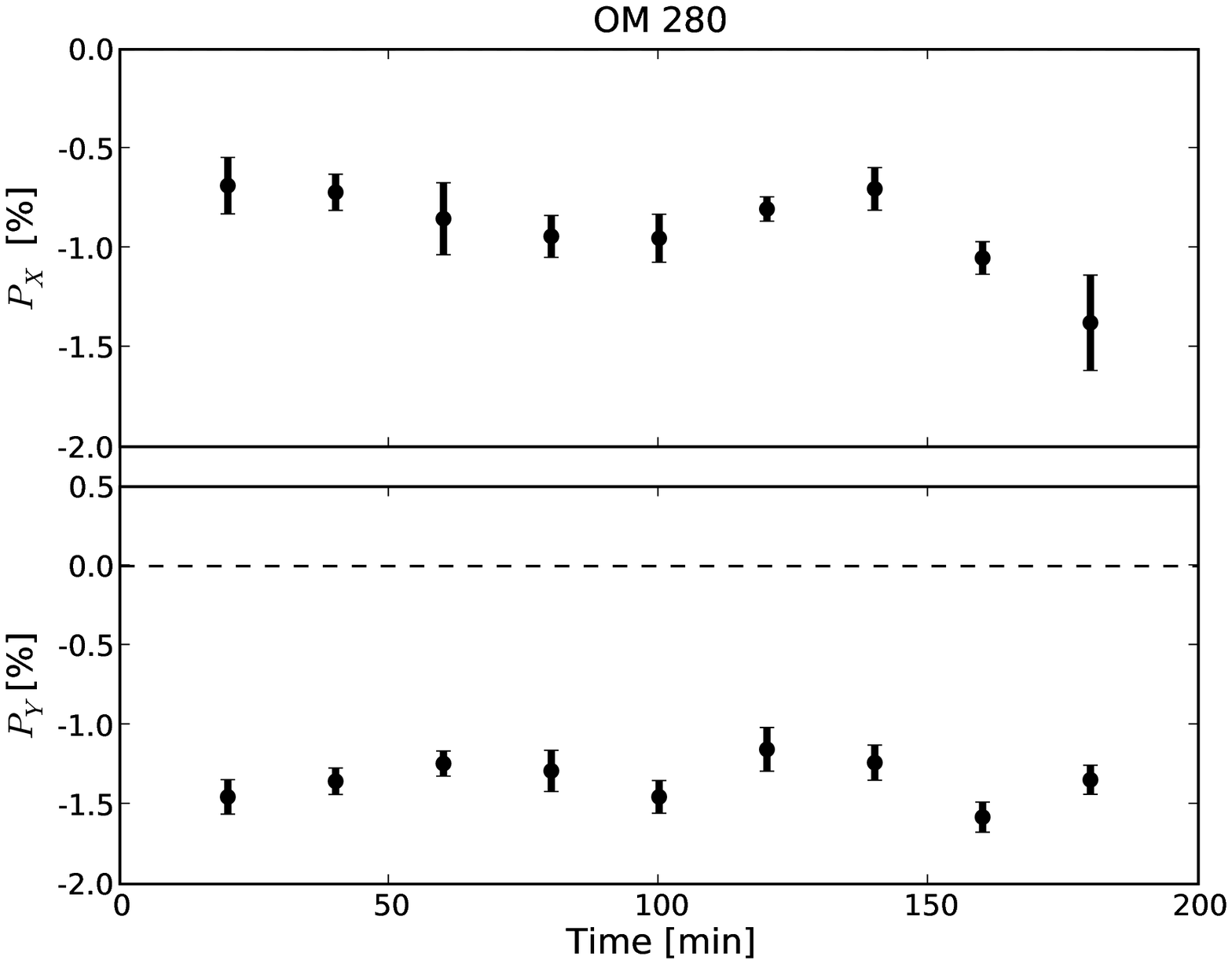}
\caption{Polarization intranight variability, continued. For caption see Fig. \ref{firstplot}.}
\end{figure}

\begin{figure}
\includegraphics[width=8cm]{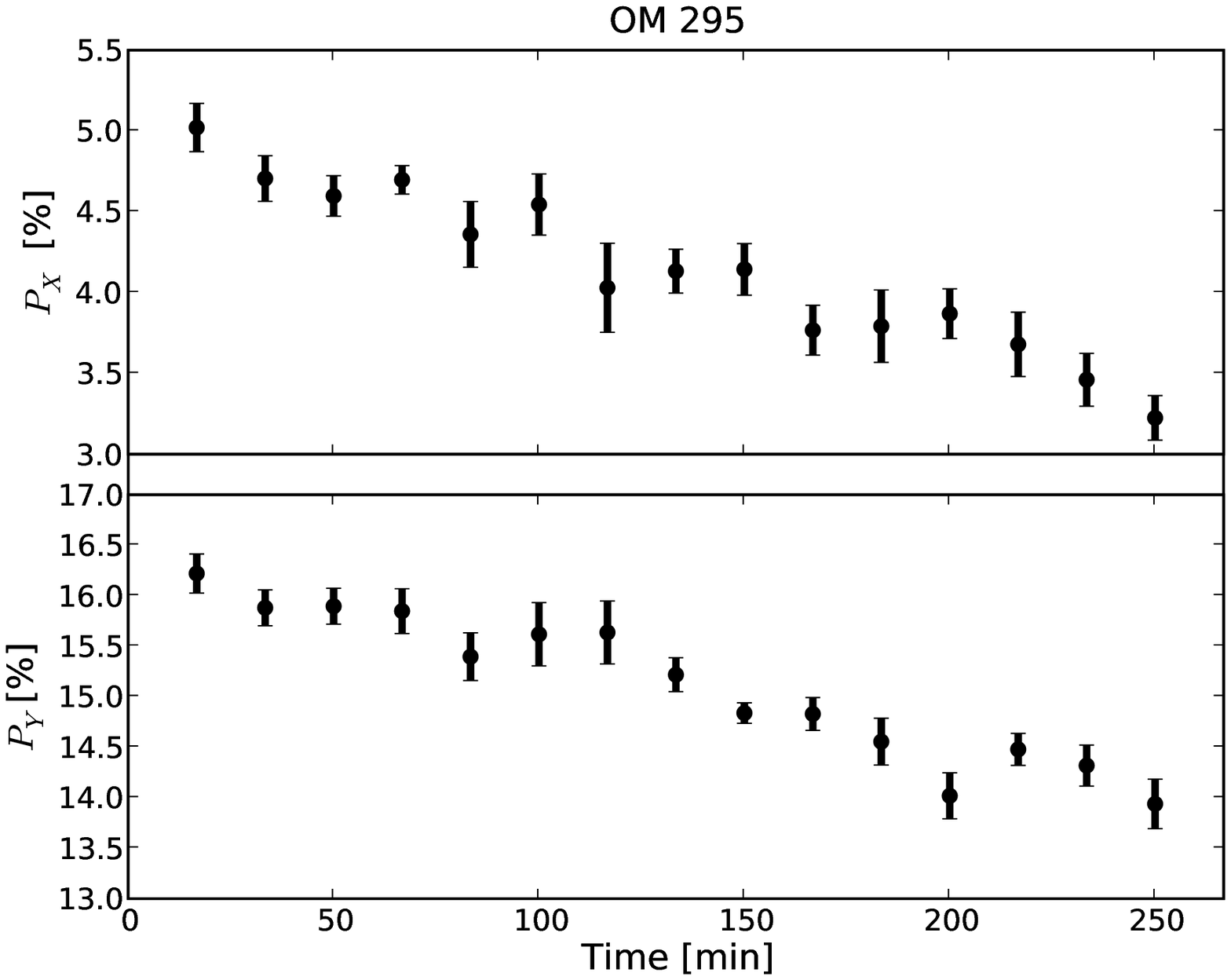}
\vspace{0.5cm}
\includegraphics[width=8cm]{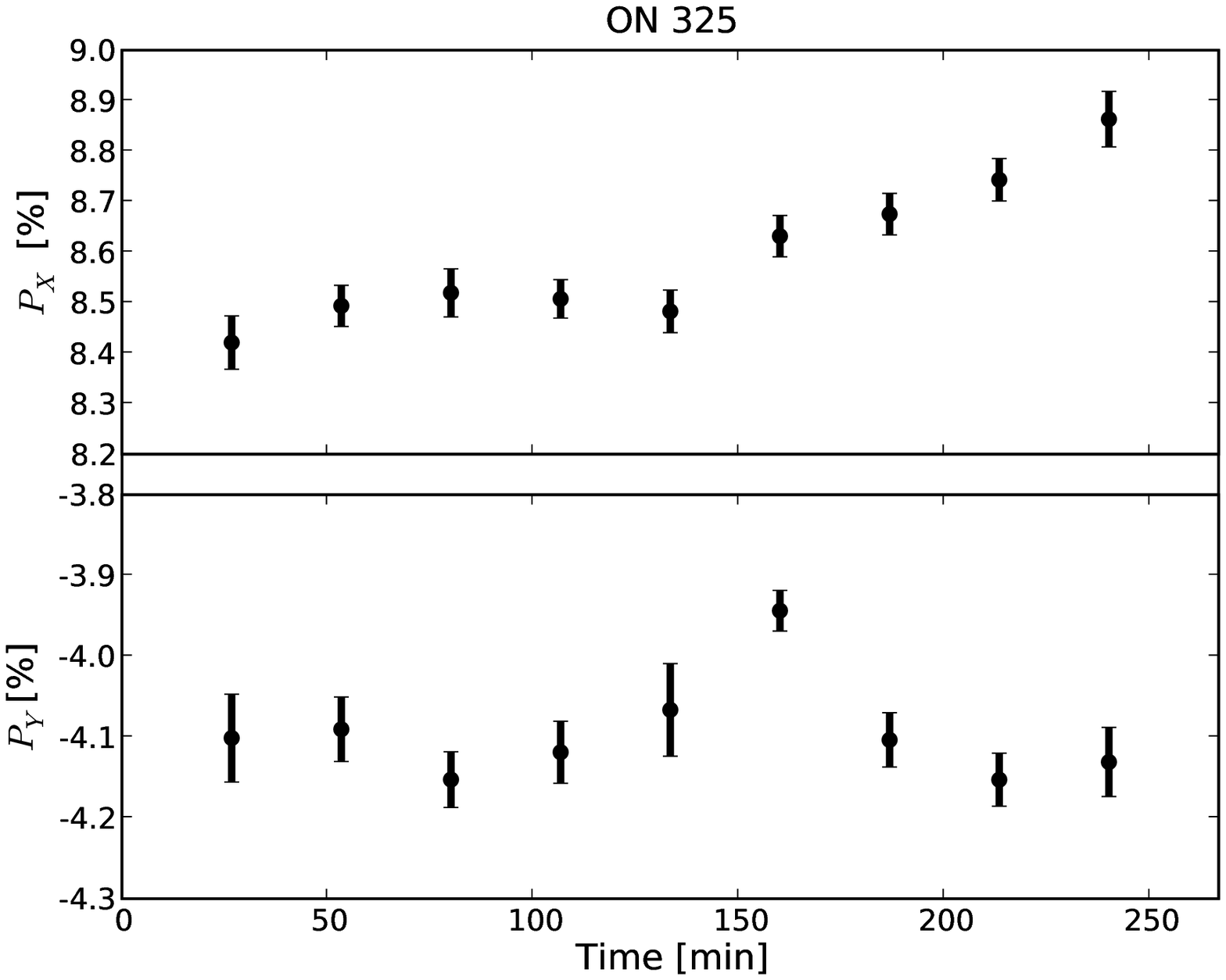}
\vspace{0.5cm}
\includegraphics[width=8cm]{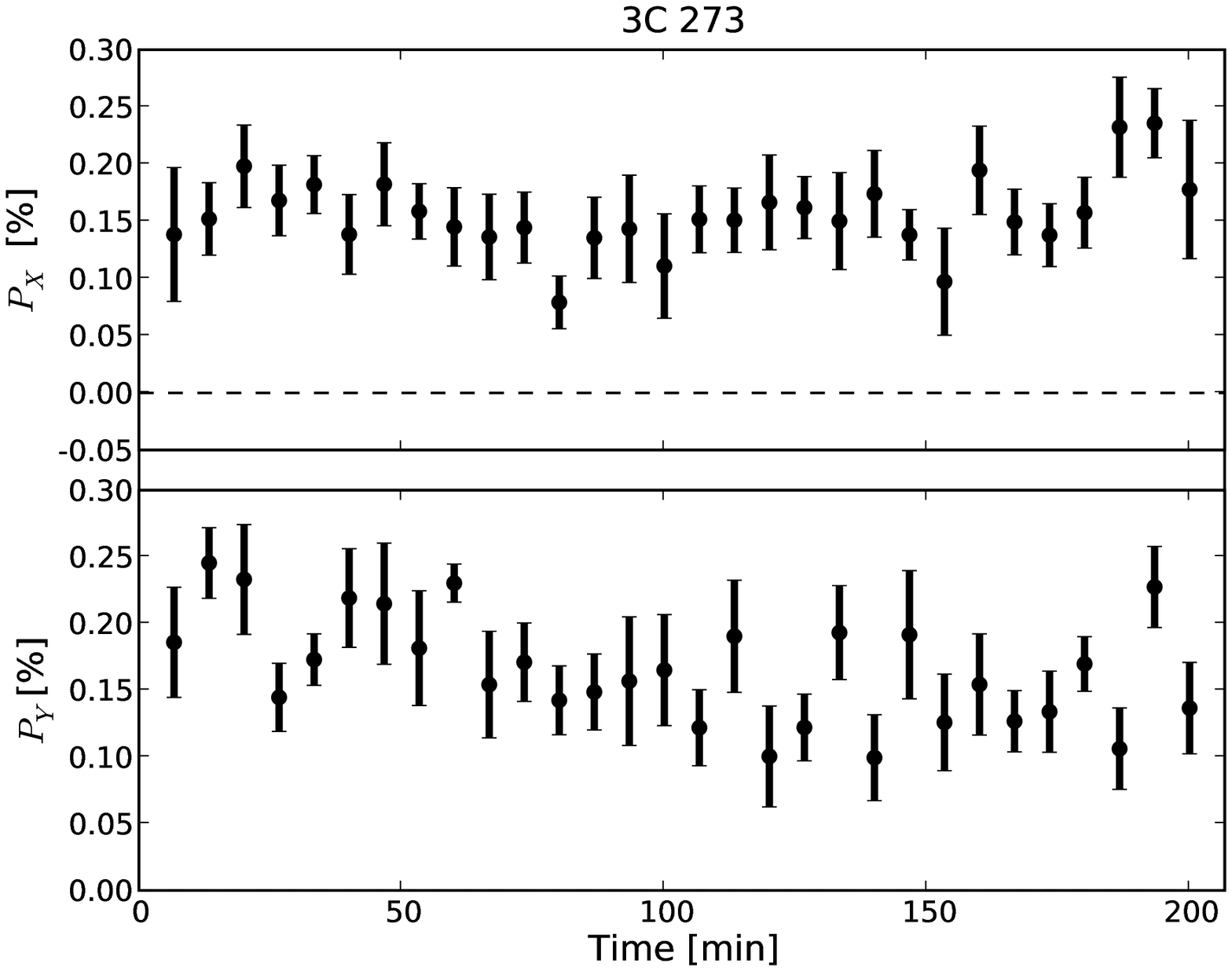}
\caption{Polarization intranight variability, continued. For caption see Fig. \ref{firstplot}.}
\end{figure}

\begin{figure}
\includegraphics[width=8cm]{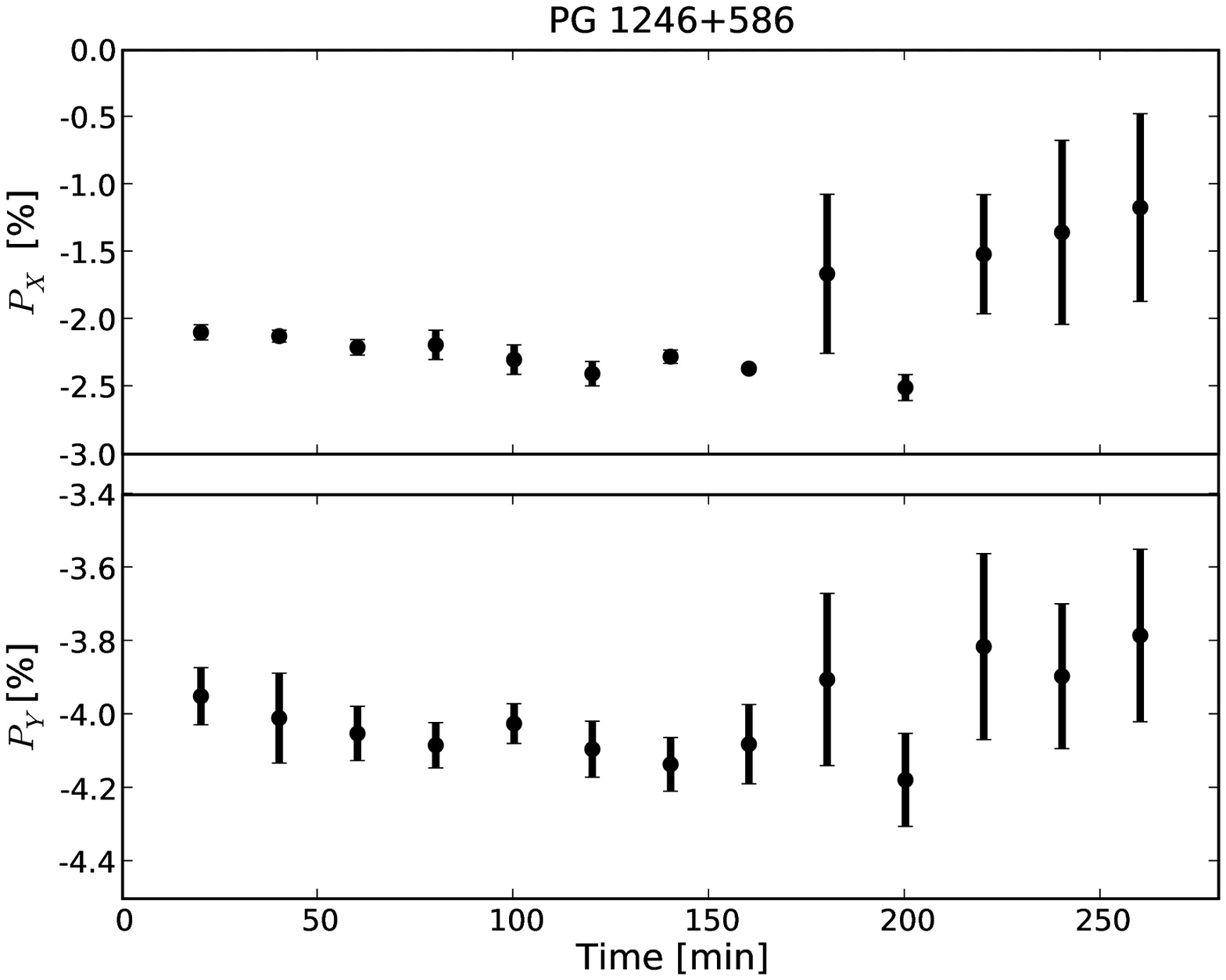}
\vspace{0.5cm}
\includegraphics[width=8cm]{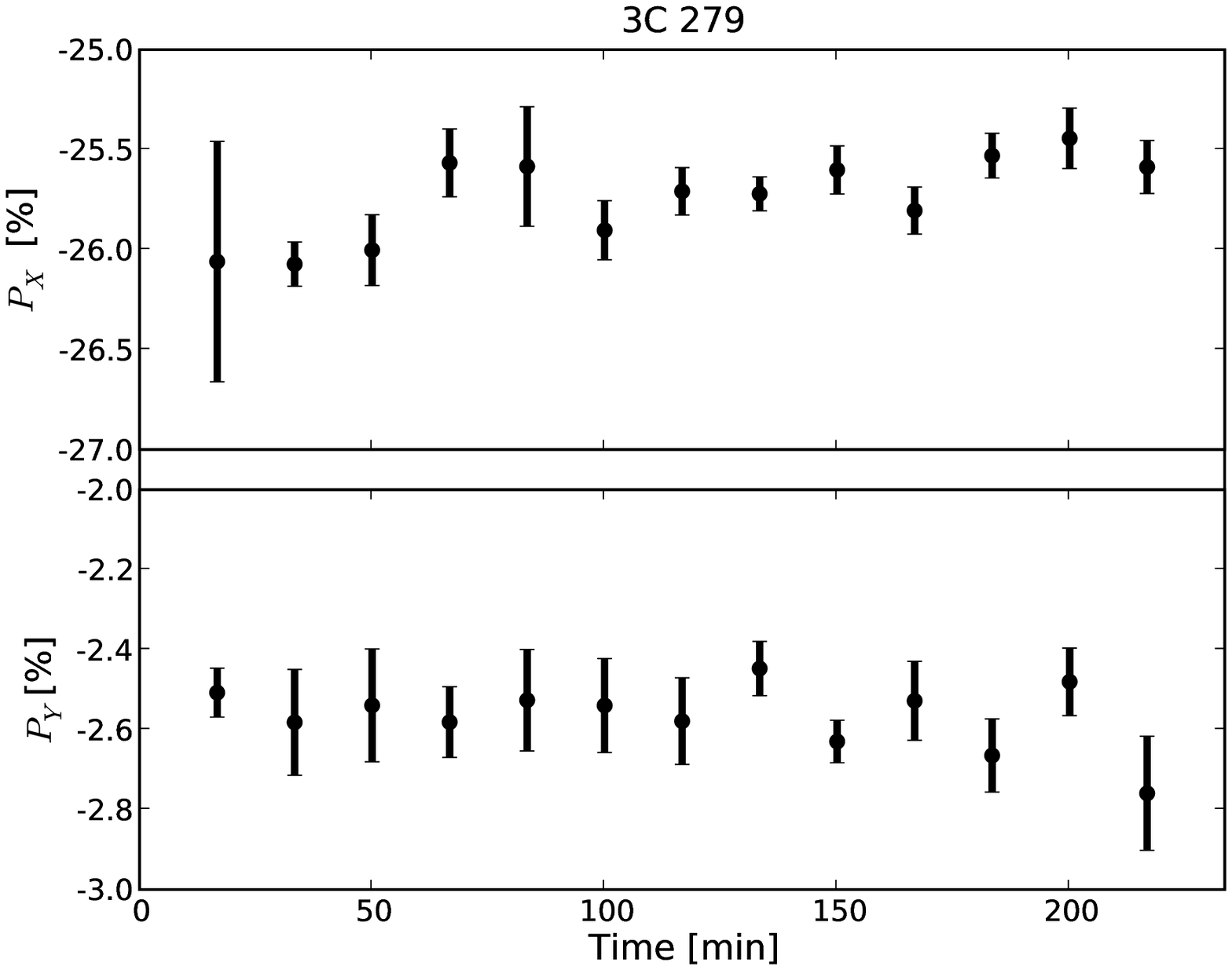}
\vspace{0.5cm}
\includegraphics[width=8cm]{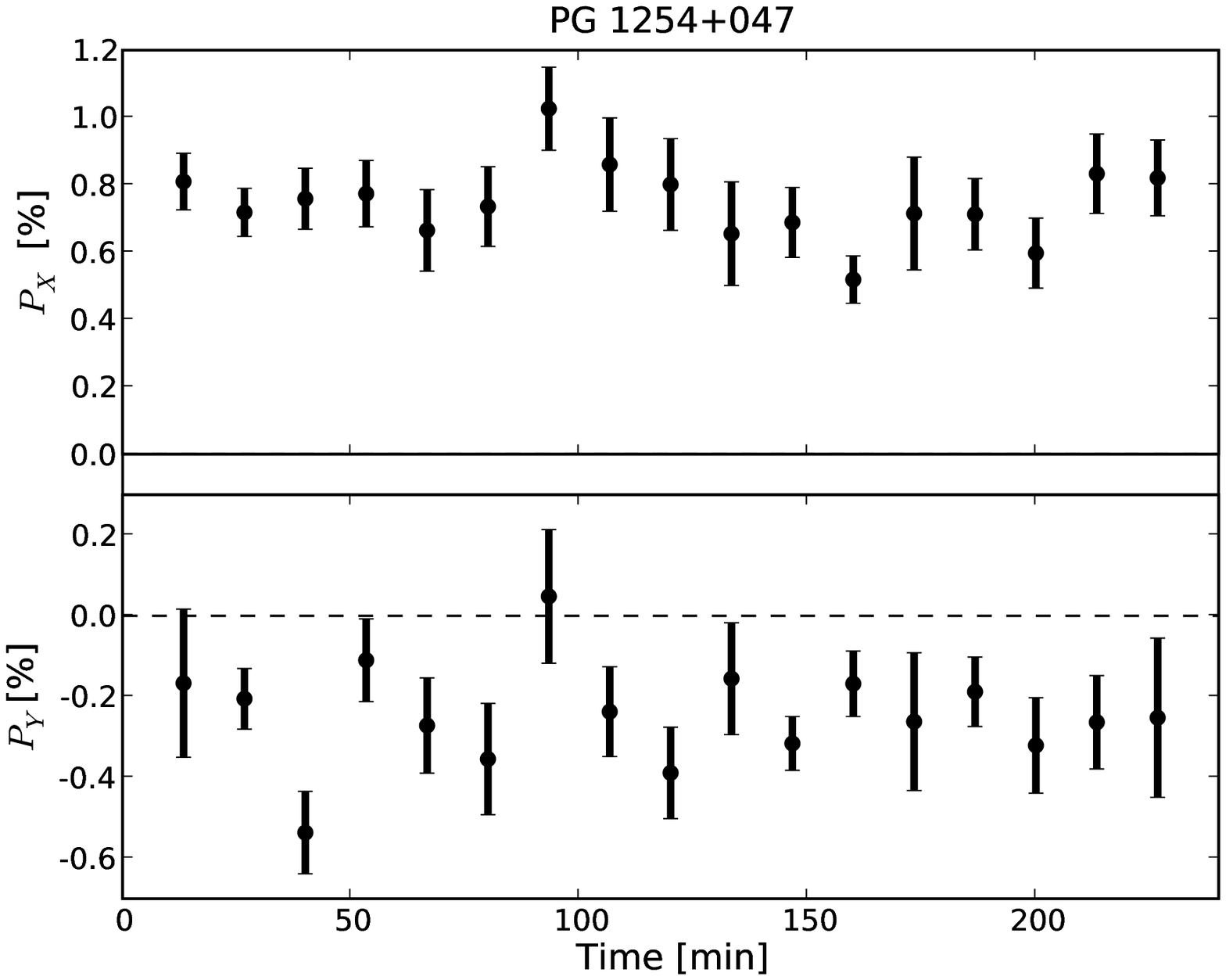}
\vspace{0.5cm}
\caption{Polarization intranight variability, continued. For caption see Fig. \ref{firstplot}.}
\end{figure}

\begin{figure}
\includegraphics[width=8cm]{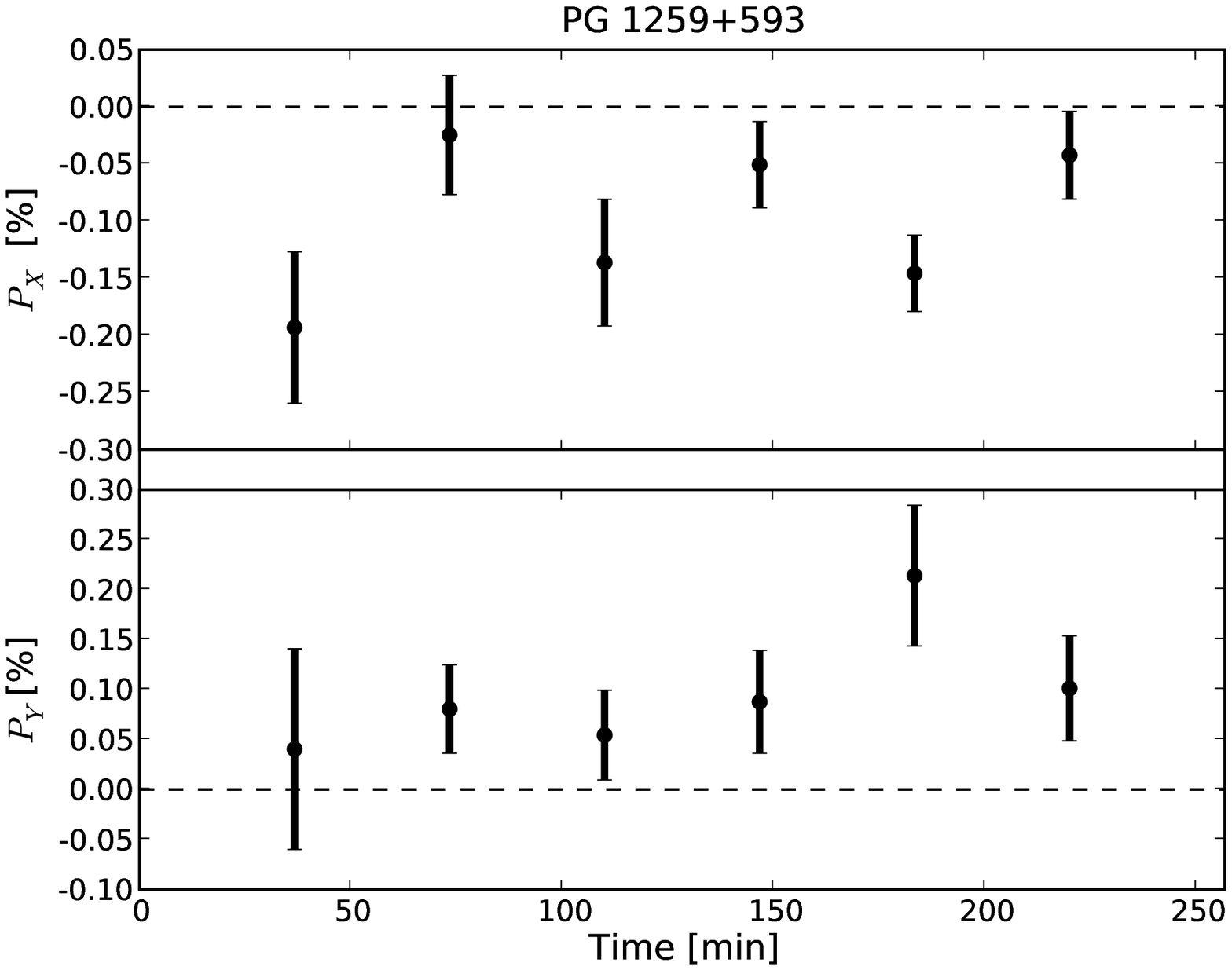}
\vspace{0.5cm}
\includegraphics[width=8cm]{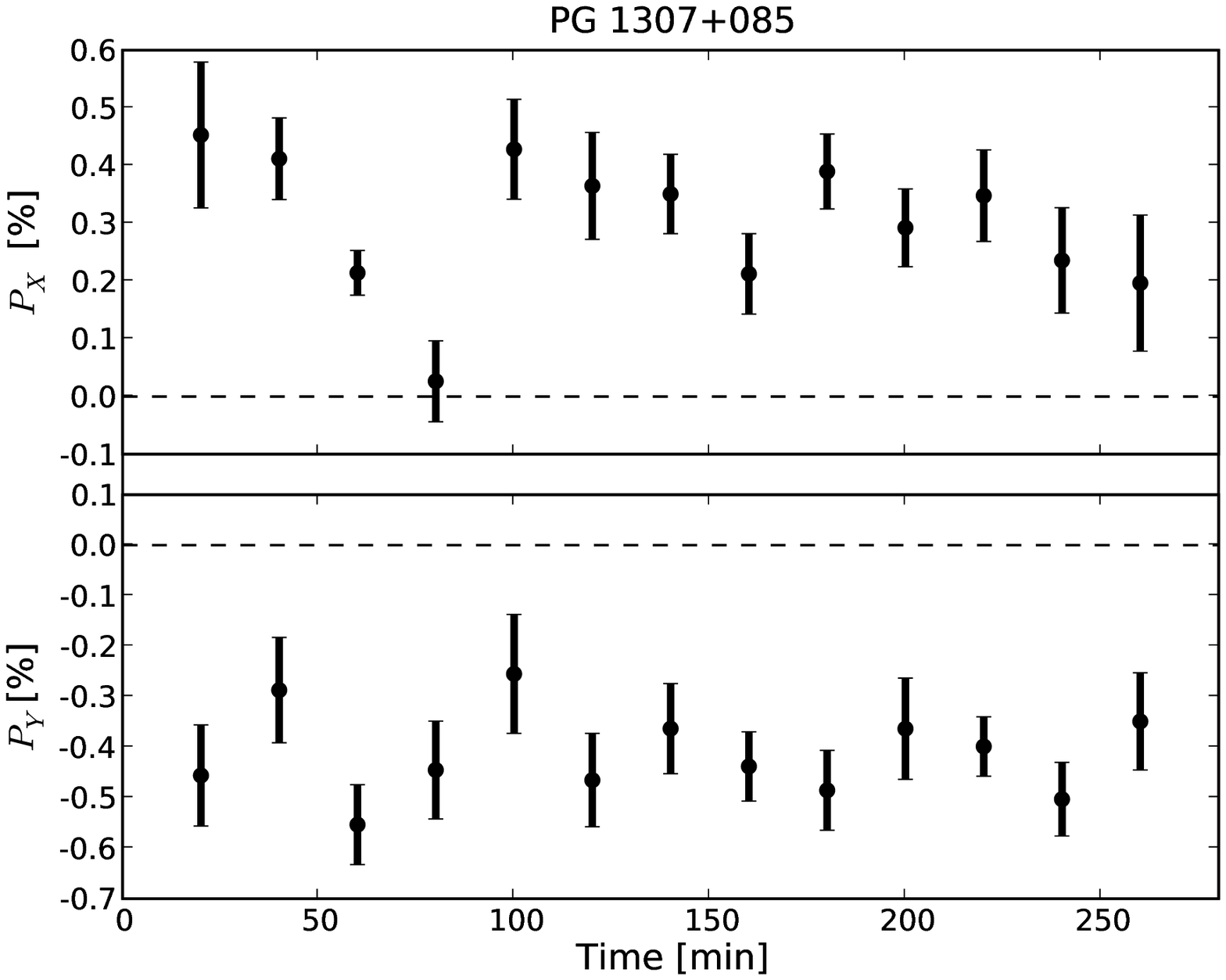}
\vspace{0.5cm}
\includegraphics[width=8cm]{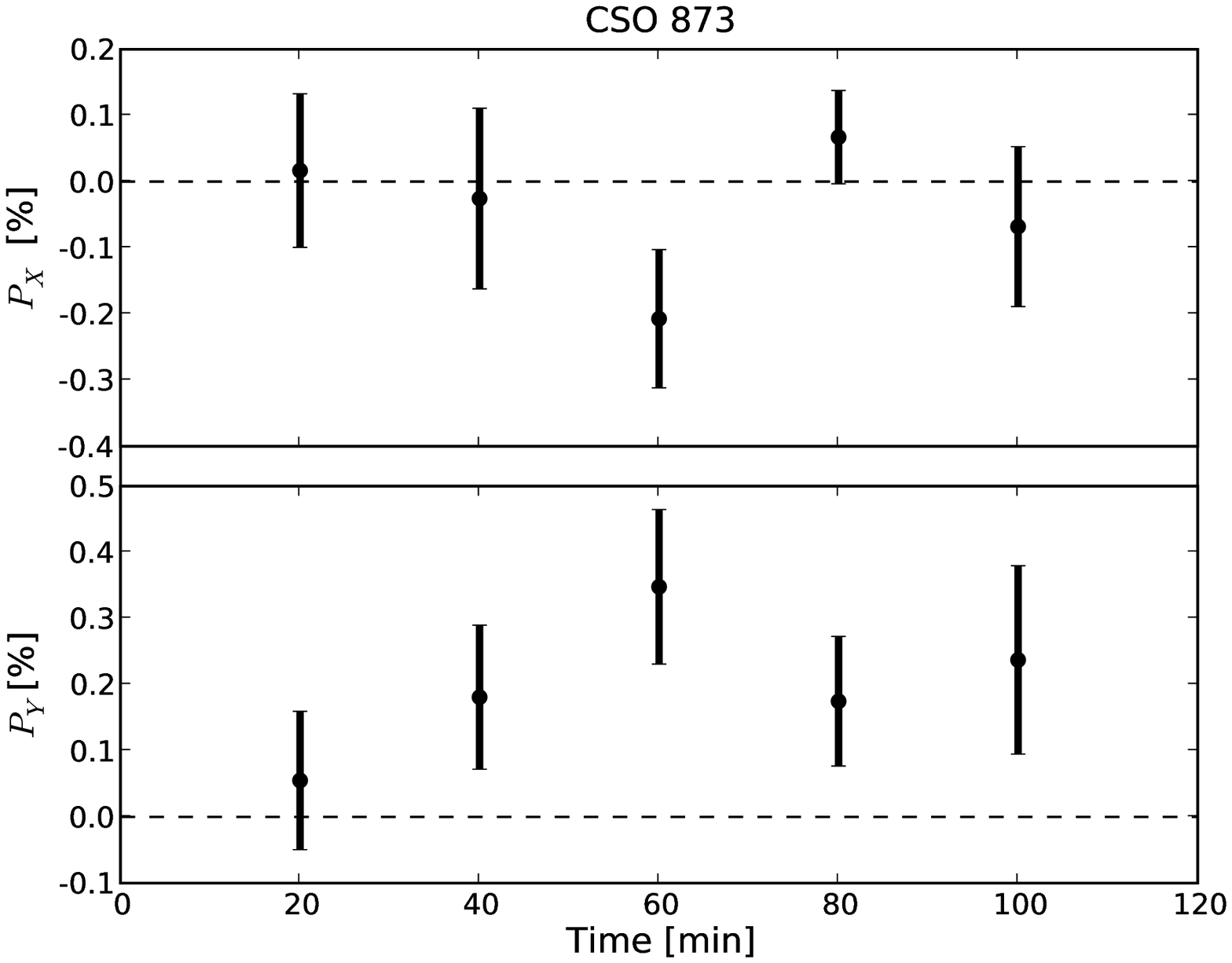}
\caption{Polarization intranight variability, continued. For caption see Fig. \ref{firstplot}.}
\end{figure}

\begin{figure}
\includegraphics[width=8cm]{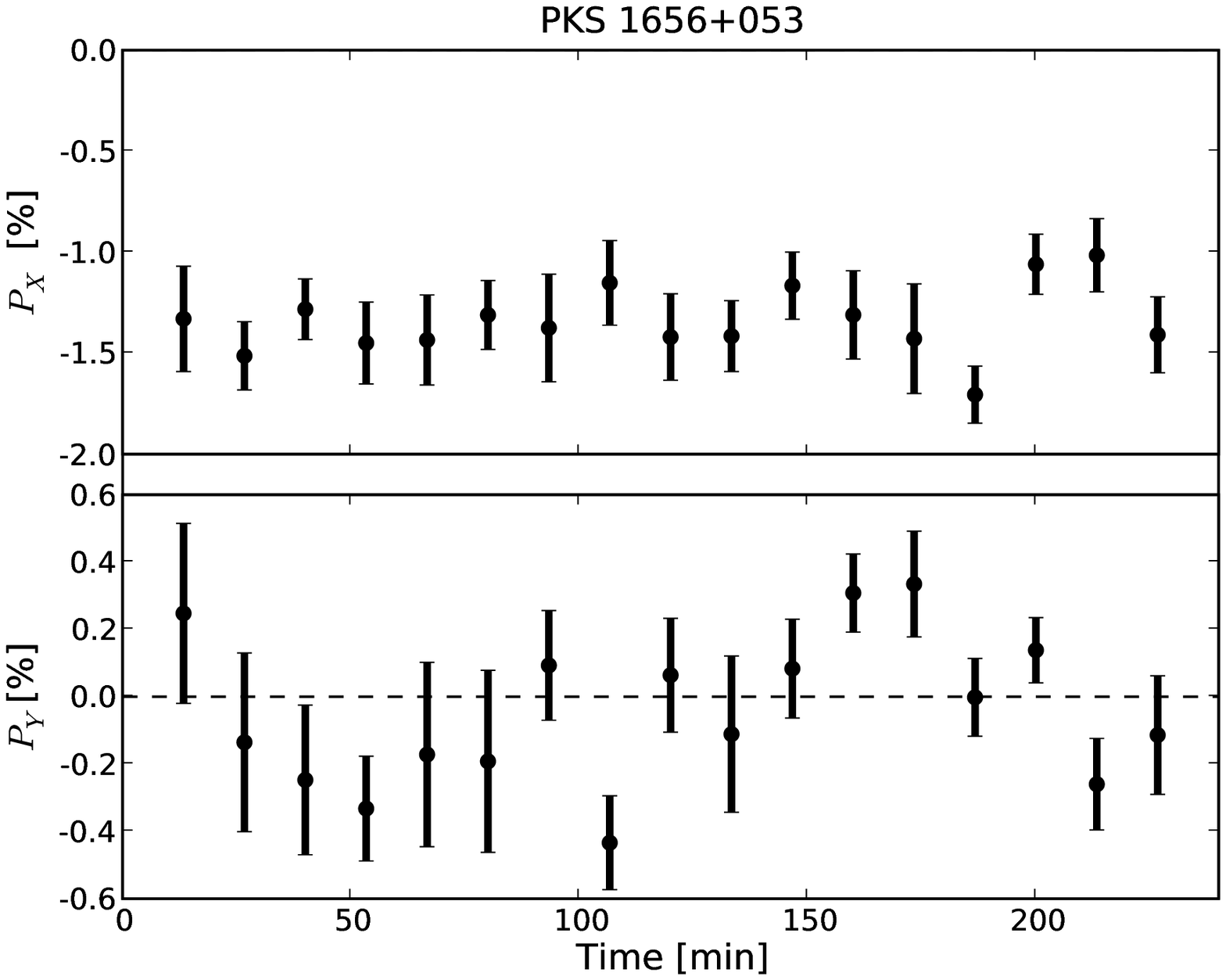}
\vspace{0.5cm}
\includegraphics[width=8cm]{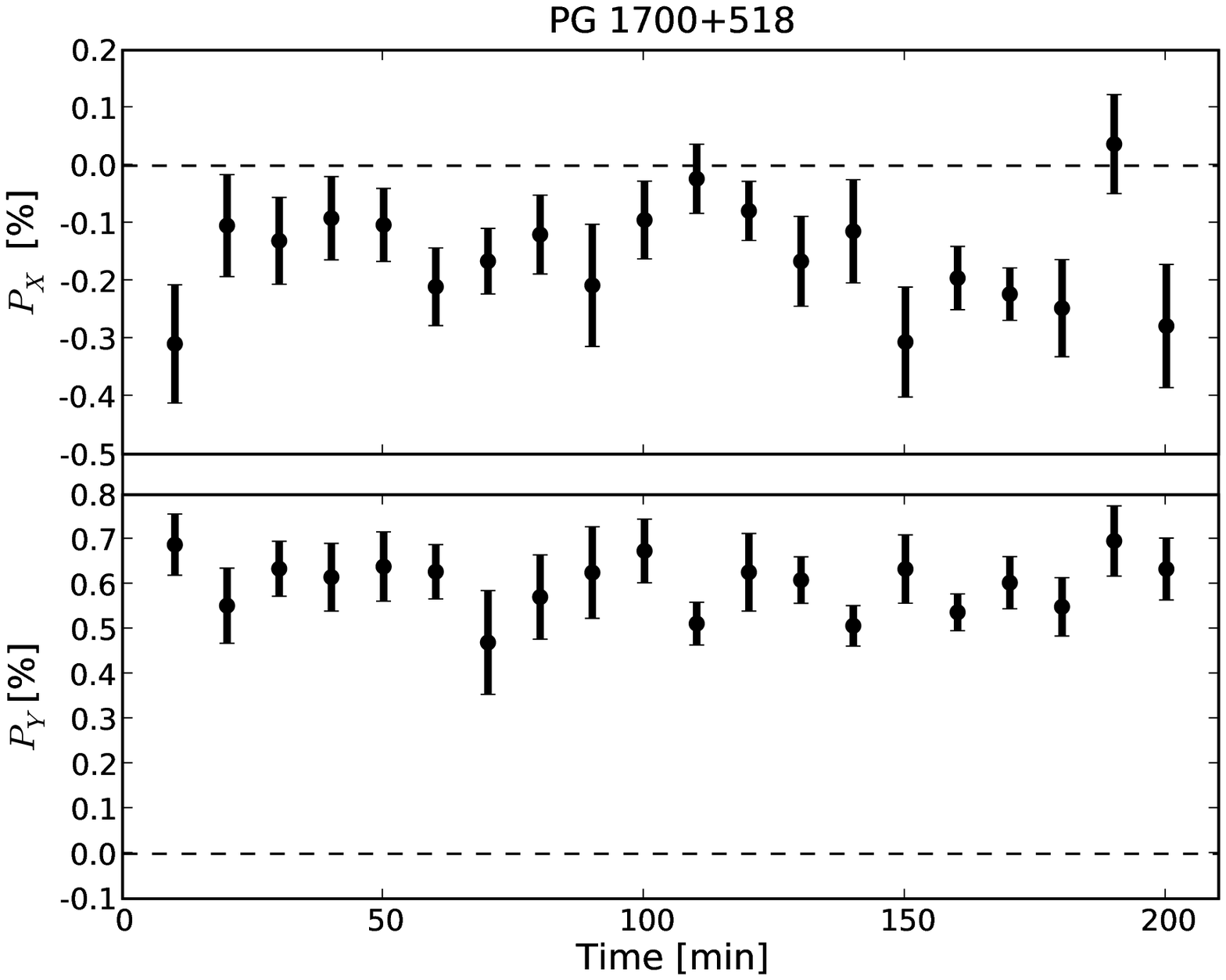}
\caption{Polarization intranight variability, continued. For caption see Fig. \ref{firstplot}.}
\end{figure}

\textbf{1ES 0647+250} is a BL Lac type object at redshift $z=0.203$. During our observations this object showed polarization of $P \sim 5$ per cent after subtraction of background polarization. However, within the four hours of observations our statistical tests as well as eyeball inspection of the data do not show variations in the degree of polarization. The object is considered non-variable in polarization.

\textbf{S5 0716+714} is a BL Lac type object at redshift $z\approx0.3$ (\citealt{Ni08}). We detect polarization of $P \sim 6$ per cent after subtraction of background polarization. One of the Stokes parameters ($P_{y}$) shows variability in form of a clear decrease of the degree of polarization. The other Stokes parameter ($P_{x}$) shows variability, however, on a much lower level.

\textbf{PKS 0735+178} is a BL Lac type object at redshift $z=0.424$. This object is polarized ($P \sim 4$ per cent) after background subtraction. Statistical tests show variability in $P_{y}$ but not in $P_{x}$.

\textbf{QSO B0754+100} is a BL Lac type object at redshift $z=0.266$. The object is highly polarized ($P \sim 11$ per cent) after background subtraction. Statistical tests imply variability in both normalised Stokes parameters. The data show a slow increase in $P_{y}$, accompanied by a simultaneous decrease in $P_{x}$. We consider the object variable in polarization.

\textbf{1ES 0806+524} is a BL Lac type object at redshift $z=0.138$. After subtraction of background polarization, we detect polarized emission ($P \sim 2$ per cent). Statistical tests imply violent variability in $P_{x}$. From eyeball inspection the variability pattern seems to show violent variability on very short timescales.

\textbf{1WGAJ0827.6+0942} is a RQQ at redshift $z=0.260$. For this object, one of the two comparison stars showed high levels of polarization ($\sim$10\%), we thus rejected this star for subtraction of background polarization. We used the other comparison star for background subtraction, this yields bigger errors than using two comparison stars. However, we can safely assume that a degree of polarization as high as measured in the first comparison star can not be due to background polarization, especially as the second comparison star shows low levels of polarization. After subtraction of background polarization, we do not detect polarized emission. The object does not show variability in polarization.

\textbf{OJ 287} is a BL Lac type object at redshift $z=0.306$. With a degree of polarization of $P \sim 16$ per cent, this is one the most highly polarized objects in our sample. Statistical tests imply variability in both normalised Stokes parameters. This result has been verified by eyeball inspection of the data. We see an decrease in $P_{y}$, accompanied by an increase in $P_{x}$. OJ287 is one of the most well observed blazars. The polarization and variability behaviour has been extensively studied by numerous groups (see e.g. \citealt{Ho84}, \citealt{Si91}, \citealt{Ta92}, \citealt{HT80}, \citealt{EVa00}). During our observations in January 2006, this object was in the decrease phase of a major outburst (\citealt{Va08}).

\textbf{FBS 0959+685} is a FSRQ at redshift $z=0.773$. After subtracting background polarization, we detect polarized emission ($P \sim 2.4$ per cent). However, the object does not show variability in polarization.

\textbf{OM 280} is a BL Lac type object at redshift $z=0.200$. After subtraction of background polarization, we detect minimal polarization of $P = 1.25_{-0.76} ^{+0.94}$ per cent. No variability in polarization is detected.

\textbf{OM 295} is a FSRQ at redshift $z=0.729$. This object shows one of the highest degrees of polarization in our sample ($P \sim 15$ per cent after background polarization subtraction). Violent variability is detected in both Stokes parameters, both show a clear decrease in degree of polarization.

\textbf{ON 325} is a FSRQ at redshift $z=0.237$. After subtracting background polarization we measure a degree of polarization of $P  \sim 10$ per cent. Statistical methods imply clear variability in both Stokes parameters. Eyeball inspection show that $P_{x}$ shows a very clear increase over the observing time, whereas the variability in $P_{y}$ shows only one \textquoteleft bump\textquoteright after about 150min of our data stream.

\textbf{3C 273} is a FSRQ at redshift $z=0.158$. After subtracting background polarization, we do not detect polarized emission in this object. Statistical tests imply variability in both Stokes parameters. Eyeball inspection verify that both $P_{x}$ and $P_{y}$ indeed seem to show variability. This might imply that the objects showed very low variable polarization that cancelled out after averaging over the whole time span of observations. We also detected variability in the polarization of both comparison stars, However, the variability in the comparison star was only caused by the first three data points, whereas the polarization variability in the object also shows when the first three data bins are removed. This seems to exclude variability in background polarization.

\textbf{PG 1246+586} is a BL Lac type object at $z=0.847$. We detect polarized emission of $P \sim 5$ per cent after background polarization. No variability in polarization is detected.

\textbf{3C 279} is a FSRQ at $z=0.536$. In our sample, this is the object that shows the highest degree of polarization with a value of $P \sim 26$ per cent after background polarization subtraction. Statistical tests imply variability in $P_{x}$, showing in an increase, whereas $P_{y}$ does not show variability.

\textbf{PG 1254+047} is a RQQ at $z=1.024$. This object was observed by \citet{GoK93}, \citet{GoK00} and \citet{Ja05} and none of those detected microvariability. After subtracting background polarization, we detect low-level polarized emission of $P = 0.85_{- 0.37} ^{+0.46}$ per cent. Please note that this might well be background polarization, in case the comparison stars are at close vicinity inside the Milky Way and the object if affected by additional background polarization. No variability in polarization is detected.

\textbf{PG 1259+593} is a RQQ at redshift $z=0.472$. After subtraction of background polarization, the measured polarization is consistent with zero polarization. Statistical test imply variability in $P_{x}$. Indeed, the polarization seems to oscillate. Variable polarization is not detected in the comparison stars.

\textbf{PG 1307+085} is a RQQ at redshift $z=0.155$. The object was observed by \citet{JM95}, they did not detect microvariability. We do not detect polarized emission. The variability detection in $P_{x}$ is most likely caused by a single outlier that is visible in the plots. As a reliability check, we remove the ten datapoints producing the outlier and rerun the ANOVA on the rest of the data. In this second ANOVA test the probability for variability dropped to $p_{var} = 89.6\%$. Thus the variability is caused by a single point and hence dubious.

\textbf{PG 1700+518} is a RQQ at redshift $z=0.292$. This object was observed by \cite{Ca07}, no microvariability was detected. After subtracting background polarization, we measure low-level polarization ($P = 0.87 _{- 0.17} ^{+ 0.21}$ per cent). However, in case of comparison stars in close vicinity inside the Milky Way, this might well be background polarization. No variability in polarization is detected.

\textbf{CTA 102} is a FSRQ at $z=1.036$. After subtracing background polarization, we measure polarized emission of $P \sim 3$ per cent in this object. Statistical tests imply variability in both $P_{x}$ and $P_{y}$, showing a \textquoteleft bumpy decrease\textquoteright of polarization over time in both values.

Overall, we observed 28 objects, 12 RQQs, eight FSRQs and eight BL Lacs. We detected polarized emission in 15/28 objects. The degree of polarization ranged from as low as $P = 0.85_{- 0.37} ^{+ 0.46}$ per cent to as high as $P \sim 26$ per cent. Two RQQs (PG1254+047, PG1700+518) showed polarized emission. For the radio-loud objects, five FSRQs and eight BL Lacs showed polarized emission. We detected variability in polarization in eleven objects. In two cases, we detect variable polarization in unpolarized objects (3C 273, PG1259+593). Nine objects are polarized and variable in polarization. All of these objects are radio-loud, four of them being FSRQs, five of them being BL Lacs. Most of the variability shows in a increase or decrease of $P_{x/y}$ over the time span of our observations. In some cases, this trend is modulated by variations on shorter timescales. We will discuss the results further in the following section. See Table \ref{duty} for a summary of the sample statistics. 

\begin{table*}
\caption{Result table. Description of columns: ID: Object name; $z$: redshift; \#: number of comparison stars observed; exp. time/cycle time: exposure time for one angle, cycle time is exposure time per angle plus read-out time; best aperture: radius of best aperture for data reduction in arc seconds; $p_{var}$ [\%]: probability for variability in $P_{x}$ and $P_{y}$ separately, meaning of flags: N=not variable, Y= variable, D=dubious, see text for discussion; $V/RV$: redshift corrected violence and relative violence of variability, see text for explanation; $P$: degree of polarization in $\%$; $PA$ corrected position angle of degree of polarization; type: object type.}
\centering
\label{resulttab}
\begin{tabular}{l c c r c r r c c c c}
\hline \hline
ID & $z$ & \# & exp. time [s] & best aperture & $p_{var}$ [\%] & $p_{var}$ [\%] & $V$/$RV$ & $P$ & PA & type \\
   &     &  & cycle time [s] & radius [''] & $P_{x}$ & $P_{y}$ &  $[10^{-3} min^{-1}]$& [\%] & [$\circ$] & (subclass) \\
\hline
PG 0026+129     & 0.145 & 2 & 15.6 (20.0) & 2.09 & 39.11 / N & 33.56 / N & - & - & - & RQQ \\
PG 0043+039     & 0.384 & 1 & 26.7 (30.0) & 1.90 & 39.65 / N & 8.74 / N & - & - & - & RQQ \\
FBQS J0242+0057 & 0.569 & 2 & 16.5 (20.0) & 1.71 & 38.84 / N & 13.38 / N & - & - & - & RQQ \\
1E 0514-0030    & 0.291 & 3 & 15.2 (20.0) & 1.52 & 39.52 / N & 77.52 / N & - & - & - & RQQ \\
1ES 0647+250    & 0.203 & 2 & 24.8 (30.0) & 3.23 & 63.03 / N & 58.33 / N & - & 4.71$\pm$0.13 & 134 & BL Lac \\
S5 0716+714     & 0.300 & 1 & 6.8 (10.0)  & 2.47 & $>$99 / Y & $>$99 / Y & 18.65/3.18 & 5.87$\pm$0.10 & 116 & BL Lac \\
PKS 0735+178    & 0.424 & 1 & 16.4 (20.0) & 1.71 & 24.57 / N & $>$99 / Y & 1.02/0.26 & 3.92$\pm$0.24 & 120 & BL Lac \\
QSO B0754+100   & 0.266 & 2 & 25.6 (30.0) & 1.33 & $>$99 / Y & $>$99 / Y & 3.99/0.37 & 10.75$\pm$0.18 & 48 & BL Lac \\
1ES 0806+524    & 0.138 & 2 & 24.7 (30.0) & 1.52 & $>$99 / Y & 47.40 / N & --/-- & 2.25$\pm$0.12 & 132 & BL Lac \\
1WGAJ0827.6+0942& 0.260 & 2 & 40.6 (45.0) & 3.04 & 83.00 / N & 39.18 / N & - & - & - & RQQ \\
OJ 287          & 0.306 & 1 & 11.7 (15.0) & 3.42 & $>$99 / Y & $>$99 / Y & 2.31/0.14 & 16.23$\pm$1.00 & 126 & BL Blac \\
QSO B0953+414   & 0.239 & 1 & 16.8 (20.0) & 1.90 & 84.87 / N & 4.58 / N & - & - & - & RQQ \\
3C 232          & 0.530 & 1 & 31.8 (35.0) & 1.52 & 34.31 / N & 26.98 / N & - & - & - & FSRQ \\
FBS 0959+685    & 0.773 & 2 & 35.7 (40.0) & 2.28 & 31.56 / N & 48.00 / N & - & 2.36$\pm$0.28 & 164 & FSRQ \\
OM 280          & 0.200 & 1 & 26.8 (30.0) & 1.90 & 49.21 / N & 35.77 / N & - & $1.25_{- 0.76} ^{+ 0.94}$ & 119 & BL Lac \\
OM 295          & 0.729 & 2 & 20.6 (25.0) & 1.52 & $>$99 / Y & $>$99 / Y & 13.14/0.85 & 15.46$\pm$0.09 & 37 & FSRQ \\
ON 325          & 0.237 & 1 & 37.4 (40.0) & 3.23 & $>$99 / Y & $>$99 / Y & 1.69/0.18 & 9.52$\pm$0.36 & 168 & FSRQ \\
3C 273          & 0.158 & 2 & 5.7 (10.0)  & 2.47 & $>$99 / Y & 95.57 / Y & - & - & - & FSRQ \\
PG 1246+586     & 0.847 & 1 & 26.7 (30.0) & 2.66 & 56.25 / N & 4.79 / N & - & 4.73$\pm$0.31 & 123 & BL Lac \\
3C 279          & 0.536 & 1 & 21.7 (25.0) & 1.90 & 98.92 / Y & 13.54 / N & 1.23/0.05 & 25.85$\pm$0.35 & 93 & FSRQ \\
PG 1254+047     & 1.024 & 2 & 15.6 (20.0) & 1.90 & 66.16 / N & 36.92 / N & - & $0.85_{- 0.37} ^{+ 0.46}$ & 170 & RQQ \\ 
PG 1259+593     & 0.472 & 2 & 50.6 (55.0) & 2.28 & 95.24 / Y & 72.12 / N & - & - & - & RQQ \\
PG 1307+085     & 0.155 & 2 & 25.4 (30.0) & 1.90 & $>$99 / D & 22.87 / N & - & - & - & RQQ \\
CSO 873         & 1.014 & 1 & 26.8 (30.0) & 1.52 & 9.93 / N & 66.97 / N & - & - & - & RQQ \\
PKS 1656+053    & 0.879 & 2 & 15.6 (20.0) & 1.52 & 48.91 / N & 88.24 / N & - & - & - & FSRQ \\
PG 1700+518     & 0.292 & 1 & 11.6 (15.0) & 1.71 & 81.54 / N & 74.42 / N & - & $0.87 _{- 0.17} ^{+ 0.21}$ & 48 & RQQ \\
PG 2112+059     & 0.457 & 2 & 56.1 (60.0) & 2.09 & 18.33 / N & 92.12 / N & - & - & - & RQQ \\
CTA 102         & 1.036 & 3 & 25.0 (30.0) & 1.33 & 96.73 / Y & $>$99 / Y & 4.81/1.87 & 2.57$\pm$0.15 & 6 & FSRQ \\
\hline
\end{tabular}
\end{table*}

\section{DISCUSSION}

In this section we will sum up and discuss our results, we will calculate duty cycles for PINOV and polarized emission for our samples. The duty cycle describes the percentage of time a system is in a certain state, i.e. in our case the time that an AGN shows polarized emission or PINOV. It should be noted though that we observed our objects for only $4$ h and not a full night. Thus, we might have missed variability or polarized emission in some objects as we did not observe them long enough. We will systematically compare the samples of radio-loud and radio-quiet AGN. We will also compare our results to those of other authors and discuss possible theoretical explanations for the observed behaviour.

\subsection{Statistical properties and violent variability}

\begin{table*}
\begin{minipage}{140mm}
\caption{Duty Cycles of polarized emission and PINOV for different samples. Percentages refer to the fraction of sources being in the respective state.}
\centering
\label{duty}
\begin{tabular}{ll|cccccc}
\hline \hline
Sample & & P:yes & P:no & P:yes,var & P:yes,not var  & P:no,var & all \\
\hline
whole sample 	& (\#) & 15 & 13 & 9  & 6  & 2 & 28 \\
whole sample 	& (\%) & 54 & 46 & 32 & 21 & 7 & \\
\hline
RL (all) 	& (\#) & 13 & 3  & 9  & 4  & 1 & 16 \\
RL (all) 	& (\%) & 81 & 19 & 56 & 25 & 6 & \\
\hline
RL (FSRQ) 	& (\#) & 5  & 3  & 4  & 1  & 1  & 8 \\
RL (FSRQ) 	& (\%) & 63 & 38 & 50 & 13 & 13 & \\
\hline
RL (BL Lac) 	& (\#) & 8   & 0  & 5  & 3  & 0 & 8 \\
RL (BL Lac) 	& (\%) & 100 & 0  & 62 & 38 & 0 & \\
\hline
RQQ 		& (\#) & 2  & 10 & 0  & 2  & 1 & 12 \\
RQQ 		& (\%) & 17 & 83 & 0  & 17 & 8 & \\
\hline
\end{tabular}
\end{minipage}
\end{table*}

It indeed seems clear from our study that radio-loud and radio-quiet AGN show distinctly different behaviour regarding intranight polarization variability. 13 of the 16 observed radio-loud objects showed polarized emission, ranging from as low as $P = 1.25\pm0.85$ per cent to as high as $P \sim 26$ per cent. The three radio-loud AGN that did not exhibit polarized emission were FSRQs. More than half of the radio-loud AGN show PINOV.

Only two of the observed 12 RQQs showed polarized emission. In both cases the measured polarization was extremely low and the detection might represent residual background polarization. Only one RQQ, PG 1259+593, shows variable polarization, however this objects is classified as unpolarized. Such a behaviour might either be explained by a flickering low level polarization or might be \textquoteleft false\textquoteright polarization variability. For example, if there is underlying constant but low polarized emission from e.g. scattering in the host galaxy or the dust torus, non-polarized varying flux from another component (e.g. the accretion disk) can cause variability in $P_{x/y}$.

The BL Lac type objects build the most extreme sample, all of them show polarized emission from $P \sim 1.3$ per cent to as much as $P \sim 16$ per cent. Five of the eight observed BL Lac type objects showed variability within an observing span of about $\sim4$ h. Three of the eight observed BL Lac type objects did not show variability in polarization. All objects with degrees of polarization greater than $5$ per cent are variable in polarization. However, also some objects showing degrees of polarization lower than $5$ per cents show variability in polarization.

The FSRQs conform a less homogeneous sample. Three of the observed eight FSRQs did not show polarized emission during the observations. On the other hand the object with the highest degree of polarization detected in this study, 3C 279 with a degree of polarization as high as $\sim26$ per cent is a FSRQ. With degrees of polarization from $P \sim 2.5$ per cent to $P \sim 26$ per cent these values span a wider range than for the BL Lac type objects. We observed one FSRQ that was unpolarized but variable in polarization (3C 273). This can be explained in the same manner as mentioned above for the RQQ showing the same behaviour.

However, the differences in the range of measured polarization between the BL Lacs and the FSRQs might well be coincidental. As the polarization is variable, we might have observed the objects in a very different state at another date.

The observed variability appears in mostly slowly increasing or decreasing $P_{x}$ or $P_{y}$, sometimes modulated with faster variability. Thus it seems that we clearly undersample the variation timescales, we can only give a lower limit for the timescales. 

\begin{figure}
\includegraphics[width=8cm]{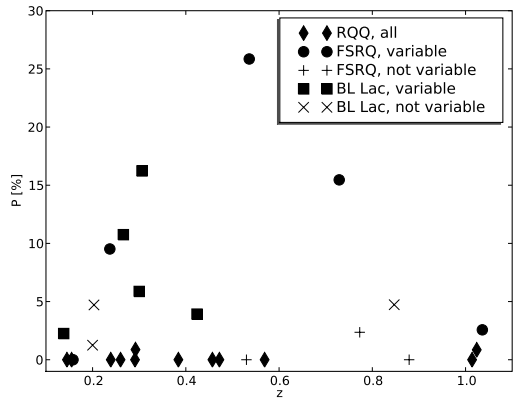}
\caption{Degree of polarization plotted against redshift for different samples.}
\label{p_vs_z}
\end{figure}

A plot of polarization variability and polarization properties against redshift can be seen in Fig. \ref{p_vs_z}. The plot shows that all objects showing degree of polarization higher than $~5$ per cent show variability in polarization. However, we detect objects with degrees of polarization lower than $5$ per cent that show polarization variability. We conclude that there might be a correlation between the degree of polarization and PINOV in the sense that all objects with $P > 5$ per cent showing variability, but not all variable objects showing degrees of polarization $> 5$ per cent.

To check the probable dependency of variability on degree of polarization, we study an estimate of the intensity of the variations: the violence $V$ and the relative violence $RV$. We define the violence of the polarization variations as follows. First, we calculate the standard deviation of the binned dataset (bin size 10) for $P_{x}$ and $P_{y}$, $\sigma_{total}$. Additionally, we estimate the measurement error as the mean standard error of the binned data (bin size 10), $S$. In principle this kind of calculation could also be done with the standard deviation of the unbinned data as $\sigma_{total}$ and the mean error of the unbinned data as $S$. Both possibilities have their advantages and disadvantages. Considering $\sigma_{total}$, this value is much better constrained for the unbinned data as the number of datapoints is ten times the number of datapoints for the binned data. On the other hand, $S$ is well defined for the binned data as it represents an empirical estimate of the error, whereas $S$ for the unbinned data is affected by the fact that we only consider shot  and read-out noise and other sources of error are not considered. We already discussed in Section 4 that systematic errors are immanent to dual beam polarimeters as used for this study (see e.g. \citealt{PR06}). We argue that the presence of systematic errors poses a more severe problem than the possibly not well constrained standard deviation when using binned data. Thus, we decide to use the binned data to estimate the violence $V$ and relative violence $RV$.

From the two values calculated as described above, we estimate the real variability range $\sigma$ in $P_{x/y}$ as:

\begin{equation}
\sigma(P_{x/y}) = \sqrt{ \sigma_{total}^{2} - S^{2} }
\end{equation}
From these values we estimate the range of the variation $r$, representing the FWHM of the distribution of the degree of polarization $P$ in polarization corrected for measurement errors as follows:

\begin{equation}
r = 2.35 \times \sqrt{\sigma(P_{x})^{2} + \sigma(P_{y})^{2}}
\end{equation}
Inserting the expressions for $\sigma(P_{x/y})$ this yields:

\begin{equation}
r = 2.35 \times \sqrt{ \sigma_{total}(P_{x})^{2} - S(P_{x})^{2} + \sigma_{total}(P_{y})^{2} - S(P_{y})^{2} }
\end{equation}
Using this value and the timespan of observations $\Delta t$ and the redshift $z$ we calculate the redshift corrected violence $V$:

\begin{equation}
V = \dfrac{r}{\Delta t} \times \left( 1 + z \right) 
\end{equation}
In the case of 1ES 0806+524, $\sigma_{total} < S$ and thus the violence $V$ is not defined. We interpret this as a sign that 1ES 0806+524 is marginally variable. As discussed above, the standard deviation $\sigma_{total}$ is not well constrained for binned data. Thus it is possible that $\sigma_{total} < S$.

\begin{figure}
\includegraphics[width=8cm]{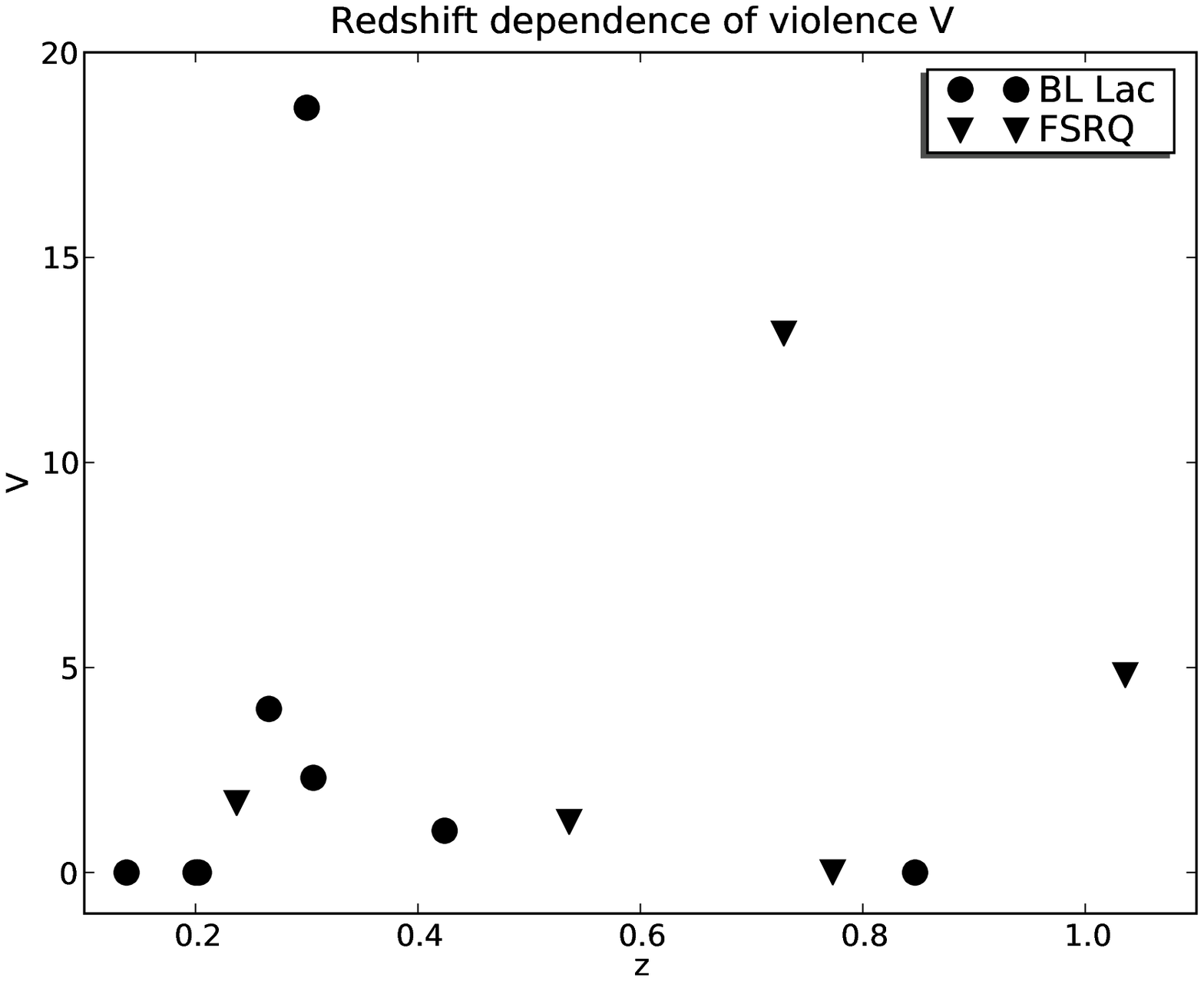}
\hspace{1cm}
\includegraphics[width=8cm]{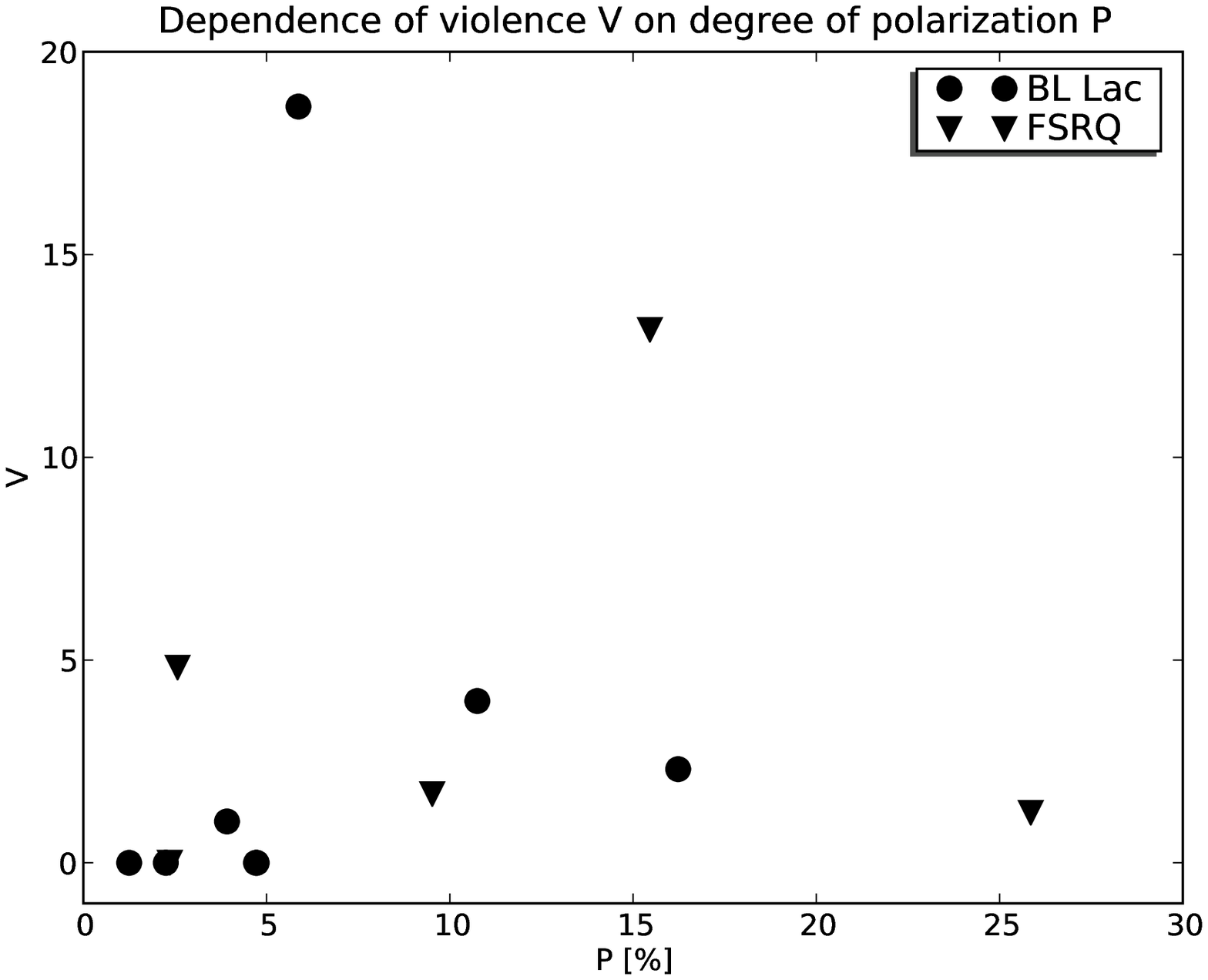}
\caption{Dependency of Violence $V$ on both redshift (Upper Panel) and degree of polarization (Lower Panel). Violence is in units of $10^{-3} $min$^{-1}$. Non-variable polarized sources are also shown in the plot, for those we set the violence to $V = 0$. However, we do not plot the values for the two possibly RQQs.}
\label{violence}
\end{figure}

We plot the violence $V$ against both the redshift $z$ and the degree of polarization $P$ in Fig. \ref{violence}. We do not find any dependency between the degree of polarization and the violence. This somewhat stands in opposition to the finding that all objects with a degree of polarization $P>5$ per cent show variability. One would expect that if variability is linked to high degrees of polarization, there would be a dependency between the intensity of variability and the degree of polarization. As we do not find such a dependency, this may either mean that low number statistics conceal possible dependencies or that the finding itself is caused by low number statistics. On the other hand it may also mean that there is a lower limit in degree of polarization and variability for a certain phenomenon. However, it is unclear what might cause such a dividing line. A possible cause for the seeming dividing line for PINOV might be the fact that detecting variability in low polarized objects is much more challenging. If we consider a highly polarized object with $P = 25$ per cent a comparatively change proportional to the degree of polarization is easily detectable, whereas the same change proportional to the degree of polarization will not be detectable with the same error range at $P \approx 1$ per cent. Bigger samples will be necessary to answer this question.

We do not see a correlation between the violence $V$ and the redshift $z$ with the violence increasing with redshift.

\begin{figure}
\includegraphics[width=8cm]{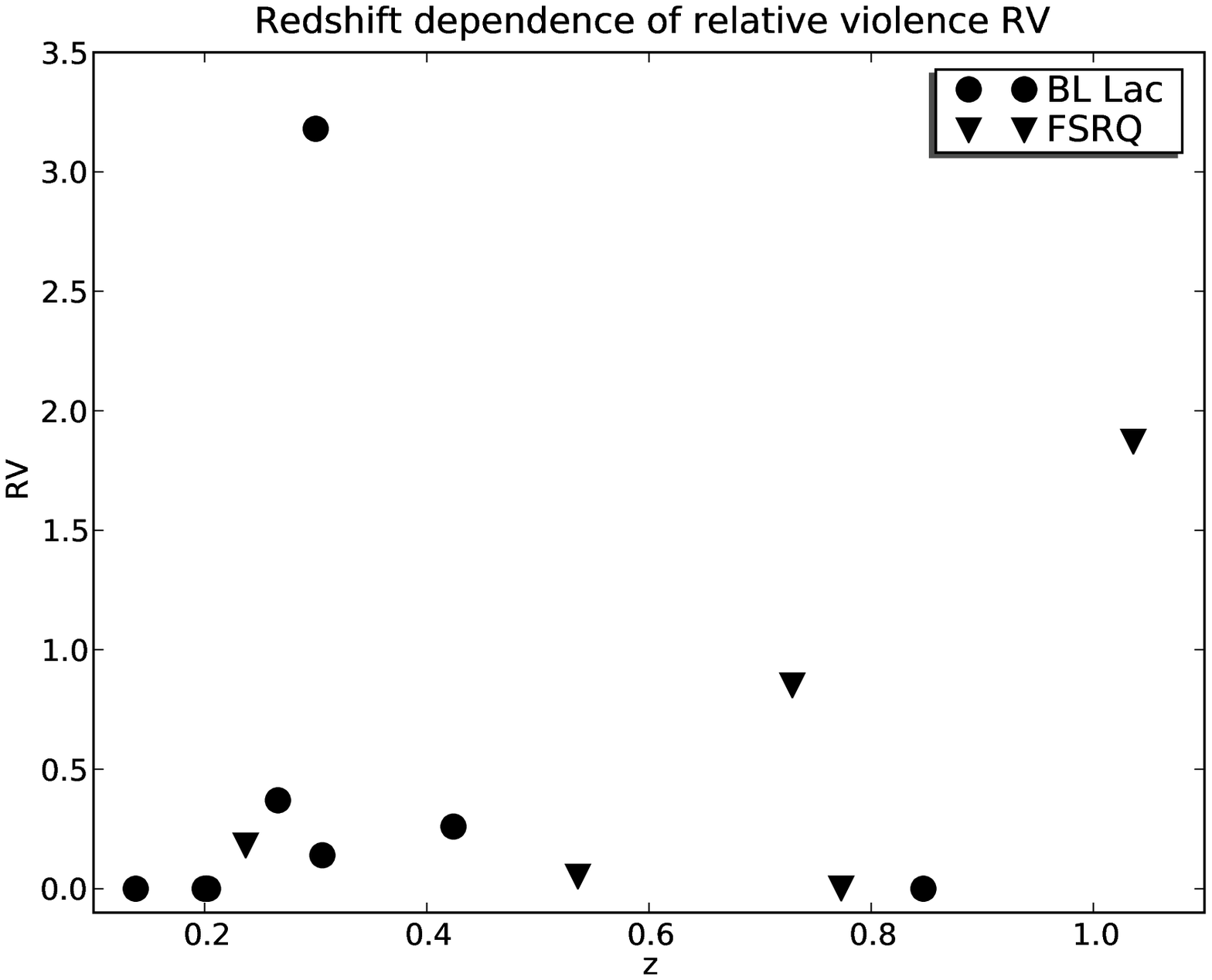}
\hspace{1cm}
\includegraphics[width=8cm]{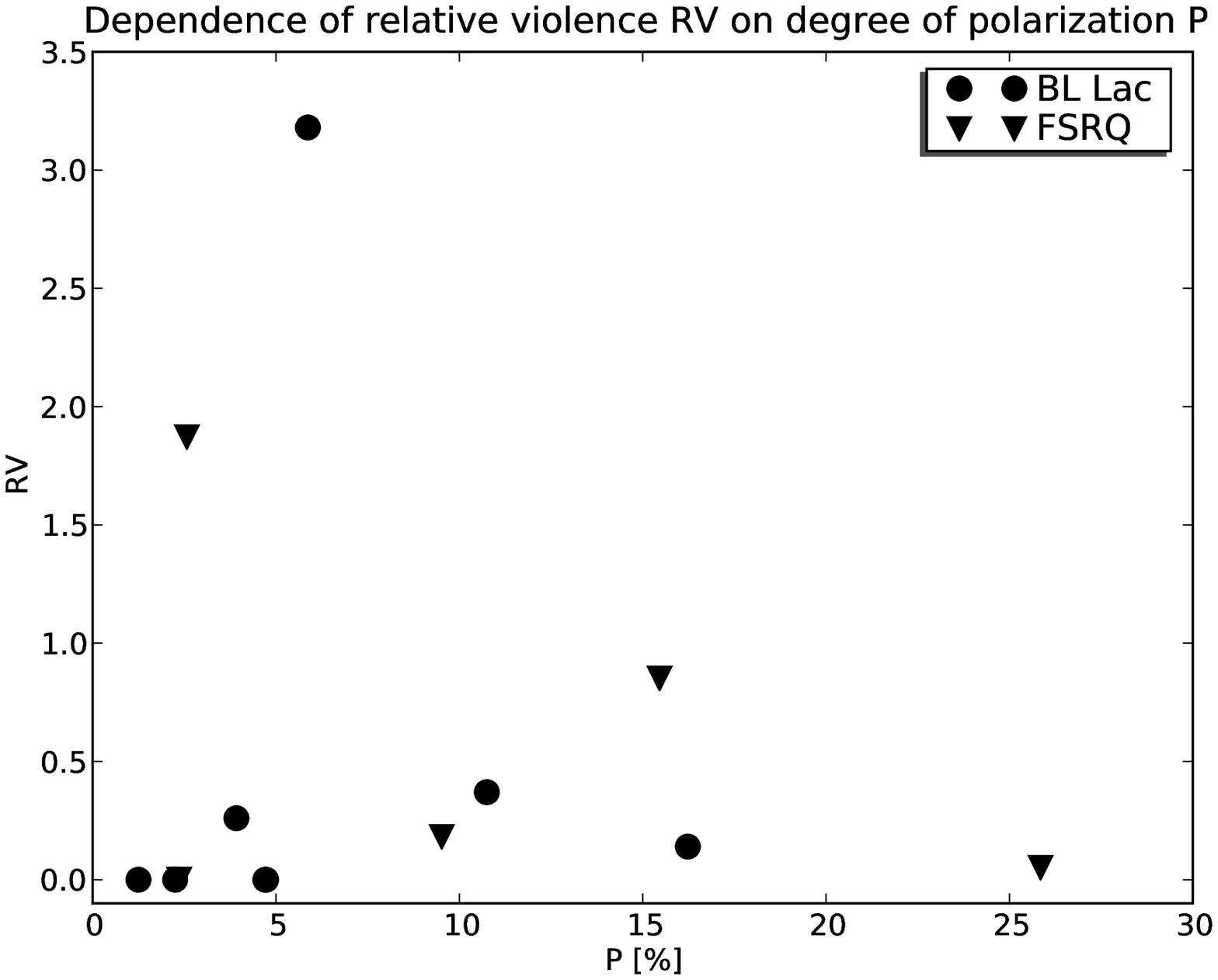}
\caption{Dependency of relative violence $RV$ on both redshift (Upper Panel) and degree of polarization (Lower Panel). Relative violence is in units of $10^{-3} $min$^{-1}$. Non-variable polarized sources are also shown in the plot, for these we set the relative violence to $RV = 0$. However, we do not plot the values for the two possibly polarized RQQs.}
\label{violence2}
\end{figure}

However, the violence might not be the best possible measure for the intensity of the variability. Thus we additionally define the relative violence $RV$ as follows:

\begin{equation}
RV = \dfrac{r}{\left\langle P \right\rangle  \times \Delta t} \times \left( 1 + z \right) 
\end{equation}
This value might be a better measure as it is normalised with the mean degree of polarization $\left\langle P \right\rangle$, calculated from the mean of the normalised Stokes parameters $P_{X/Y}$. Whilst the violence would yield the same value for a change from $P = 25$--$26$ per cent and a change from $P = 1$--$2$ per cent, the relative violence considers the comparative change in degree of polarization. However, calculating the relative violence $RV$ in this way raises the problem that the correlation between $RV$ and $P$ is somewhat implicit as $RV$ is calculated using $P$. We still show the same plots as for the violence $V$ in Fig. \ref{violence2}.

\begin{table*}
\begin{minipage}{60mm}
\caption{Spearman rank-order correlation coefficient $\rho$ and corresponding $p$-value (in brackets) for different combinations of the variables violence $V$, relative violence $RV$, degree of polarization $P$ and redshift $z$.}
\label{spearman}
\begin{tabular}{c|cc}
\hline \hline
      &   $P$          &       $z$       \\
\hline
$V$   &  -0.17 (0.69)  &   0.12 (0.78)   \\
$RV$  &  -0.67 (0.07)  &   0.17  (0.69)  \\
\hline
\end{tabular}
\end{minipage}
\end{table*}

As expected we see a tendency in the correlation between the relative violence $RV$ and the degree of polarization, with higher degrees of polarization showing lower relative violence. As discussed, this tendency might be implicit as $RV$ is calculated using $P$.

There seems to be a weak positive correlation between the relative violence $RV$ and the redshift. Still, with only nine datapoints we are clearly limited by small number statistics. A possible dependency on redshift might be caused by three different conditions. On the one hand there might be a real redshift dependency of variability intensity, on the other hand, the observed effect might be caused by the fact that we observe different restframe wavelength ranges due to the different redshifts of the objects. This would mean that variability is more violent for bluer restframe wavelength. However, objects at higher redshift have to be intrinsically brighter so we might be observing a luminosity effect.

To check for possible correlations between the violence $V$, relative violence $RV$, degree of polarization $P$ and redshift $z$ we calculate the Spearman rank-order correlation coefficient for the data plotted in Fig. \ref{violence} and Fig. \ref{violence2}. The Spearman rank-order correlation coefficient is widely used in statistics to test how well an arbitrary monotone function describes the relation between two variables. We present the results of the Spearman test in Table \ref{spearman}.

We calculate the Spearman-rank correlation coefficient $\rho$ and the corresponding $p$-value using the statistic package of SciPy\footnote{www.scipy.org}. $p$ denotes the probability that the measured correlation coefficient or a higher correlation coefficient originates from random data. However, it should be noted that  $p$ can not be calculated reliably for small datasets with $<$500 data points, we still calculate $p$, but the reader should keep in my mind that the values merely represents estimates.

For the relations between the variables $V$ and $z$, $V$ and $P$, $RV$ and $z$, the Spearman rank-order test indicates small correlations. Correlations of both violence parameters and the redshift are positive, whereas the correlations between the violence parameters and the degree of polarization are negative. However, as stated above, we are clearly limited by small number statistics. This fact also shows in the $p$-values, for the relations $z$--$V$, $z$--$RV$ and $P$--$RV$ the probability that the measured weak correlation is caused by random data is $>$ 50 per cent. Thus we can safely say that we did not detect any correlations for these values. For the parameters $RV$ and $P$, the Spearman test indicates a strong negative correlation, i.e. for higher values of $P$, $RV$ is smaller. The probability that the correlation coefficient is caused by random data is $~$7 per cent. Thus the correlation is not significant, however, one should keep in mind that we are clearly limited by low number statistics. As discussed above, this correlation is somewhat intrinsic as $P$ is used for calculating $RV$.

Further observations will have to show if the possible correlations between different parameters are spurious.

Differences between the samples of FSRQs and BL Lac type objects are not significant due to the extremely small samples. Both the violence and the relative violence show similar ranges for the two samples. Different dependencies on redshift are not detectable as the redshift distributions of the two samples are different.

\subsection{Comparison of our results with those of other authors}

PINOV has not been studied systematically so far. However, several groups studied INOV and microvariability in RQQs and RLQs and tried to study if they show different behaviour concerning INOV. In this section we will thus sum up studies about microvariability and intranight variability in quasars so far and discuss the relations between the observed INOV and PINOV properties of different types of AGN.

\citet{GoK93} studied two RQQs searching for microvariability but did not detect any signs of the very.

\citet{deD98} studied intranight variability in a well-selected sample of 34 quasars, half of them RQQs, half of them core-dominated RLQs. They found five out of 17 RQQ to show variability, for the RLQs, depending on the determination of variability, $5$--$8$ sources out of 17 showed variability. Using elaborate statistical tests, \citet{deD98} did not find significant differences in the observed variability between the two subsamples. This might imply that INOV is not mainly dependent on radio-loudness.

\citet{GoK00} observed 16 RQQs and one radio-weak quasar searching for microvariability. Determination of duty cycles is difficult for their sample as they used a wide variety of flags for the detection and non-detection of variability, owned to the fact that they tried to specify the variability patterns observed.

\citet{St05} observed 20 high-luminosity QSOs, 12 of which were RQQs. INOV was detected in more than half of the objects. Collecting data from the literature they found that opposite to BL Lacs, RLQs and RQQs show similar and comparatively low duty cycles for INOV. This seems to imply that INOV is not mainly dependent on radio-loudness, but mostly on the orientation of the relativistic jet.

\citet{Ca07} observed a sample of seven quasars, five of them RQQs, two of them radio-intermediate quasars and did not detect intranight variability in any of them. They also collected data from the literature and found radio-loud objects to show much higher percentage of intranight variability than radio-quiet objects. However, contrary to other studies this sample is not well selected and different observing strategies make a clear decision about duty cycles impossible. 

\citet{Go07} monitored 11 RQQs over 19 nights for intranight variability and calculated a duty cycle of $\sim 8$--$19$ per cent using the same statistical limits for detection of variability advocated by \citet{deD98}.

A main problem comparing the different studies is the fact that different authors use different methods to decide if an object is variable or not. All kind of methods from ANOVA (\citealt{deD98}) over simple $\chi^{2}$-methods (\citealt{Ja05}) to \textquoteleft eye-ball decisions\textquoteright (\citealt{GoK00}) were used, producing a big amount of data that is difficult to compare.

\citet{An05} studied PINOV in a sample of 18 BL Lac type objects (eight of them X-ray selected, ten radio-selected). To our knowledge this is the only systematic study of PINOV in AGN published. They found high duty cycles for variability in both the degree of polarization and the position angle. Most of the objects showed moderate to high degrees of polarization. However, their results can hardly be compared to our paper as a $\chi^{2}$-criterion was used by \citet{An05} to decide about the variability. \citet{PR06} presented a study in which systematic errors for dual beam polarimeters were found. These systematic errors are also known for ALFOSC \footnote{see http://www.not.iac.es/instruments/alfosc/polarimetry/accuracy.html} and are immanent in dual beam polarimeters as used also by \citet{An05}. Thus we suspect that part of the detected variability in this study may be caused by these systematic errors.

As \citet{An05}, we find a extremely high duty cycle for polarized emission in BL Lacs. This is however not a surprise since polarized emission is one of the defining criteria of BL Lac objects. As for the duty cycles for PINOV in BL Lac type objects, we find similarly high duty cycles as in \citet{An05}. Due to low number statistics the expected differences caused by the use of different statistical methods might not show.

Studying not only microvariability but also polarization in this context clearly breaks the degeneracy between RQQs and RLQs. Their polarization properties are clearly distinct. This indicates that the emission mechanisms in RQQs and RLQs are clearly different.

\subsection{Consistency with theoretical models}

Different theoretical models exist that explain variability on different timescales. In this section we will present these models and discuss them with respect to our results.

\citet{Cz06} discussed the role of the accretion disk in AGN variability for a geometrically-thin, optically thick disk. The dynamical time scale for such a disk can be calculated as follows:

\begin{equation}
t_{dyn} = \sqrt{\dfrac{G M}{r^{3}}}
\end{equation}
With a time scale of $~4$ h and a black hole mass of $M_{BH}=10^{9} \times $M$_{\odot}$ this yields a radius of $r<R_{Schwarzschild}$. Thus dynamical instabilities can not explain variability on intranight timescales.

The thermal time scale can be calculated as:

\begin{equation}
t_{thermal} = \alpha^{-1} \times t_{dyn}
\end{equation}
with $\alpha$ being the disk parameter described by \citet{SS73}. Assuming a characteristic value of $\alpha = 0.1$, this again yields $r<R_{Schwarzschild}$ and thus thermal instabilities can not cause variability on intranight timescales.

The viscous timescale can be calculated as:

\begin{equation}
t_{viscous} = t_{thermal} \times \left( \dfrac{r}{h_{d}} \right) ^{2}
\end{equation}
with $h_{d}$ being the disk thickness. So depending on the thickness of the disk, the viscous timescale could exhibit the right timescales to explain the intranight variability.

Most theoretical models sadly lack timescale estimations, however, combinging theoretical models (see e.g. \citealt{MW93}, \citealt{LC08}) with the time estimates from \citet{Cz06} makes accretion disc instabilities quite unlikely on time scales of a few hours, only viscous instabilities can happen on reasonable radii for the given black hole mass and time scale. Thus it can be stated that early assumptions by e.g. \citet{GoK93} that accretion disk instabilities are observed as microvariability in RQQs, would point towards viscous disk instabilities.

We interpret our results as follows: The dependency between high levels of polarization and PINOV seems to indicate that PINOV is a jet phenomenon as highly polarized emission is not expected from the accretion disk. This implies that physical conditions in the jet that cause PINOV have a very high duty cycle. Whatever causes these variations should be different from shock-in-jet phenomena as described by \citet{MG85} and observed e.g. by \citet{Ma08} as these shock fronts in jets evolve on timescales of weeks or months. However, turbulence in the shocks could explain PINOV and INOV. Changes in polarization properties in blazars have been observed by e.g. \citet{Ho84} on timescales of several days. These kind of changes should be visible on intranight scale. However, physical conditions behind these observations are still unclear. Multiwavelength studies of such variations might be a way to distinguish between different models. Observations of this phenomenon so far have been described by several variable components in the jet with different magnetic field orientations and strengths (see e.g. \citealt{Ho84}, \citealt{Ta92}). However, this is not a physical description. Theoretical models exist that can explain relatively fast variation in polarization properties (see e.g. \citealt{BB82}, \citealt{CIB85}) but have not been tested so far.

Assuming jet emission as a main cause of PINOV implies that the jet emission in RQQs is either very weak or dominated by other regions of the jet than in radio-loud AGN. The low levels of polarization observed in this study might be due to residual background polarization or originate in scattering phenomena in the accretion disk or surrounding dust, however, this polarization does not seem to change on small time scales. The orientation of the jet also affects the brightness, variability and polarization properties. The smaller the viewing angle the more the radiation from shocks is beamed and thus the degree of polarization is higher, the objects seems brighter and variability appears to be faster.

Sadly, we were not able to constrain flux intranight variability in our study, thus the link between flux and polarization intranight variability is still unclear. However, as radio-loud AGN show high duty cycles in both flux and polarization, they must coincide in many cases. Additionally, as we do not detect polarization variability in RQQs expect in one questionable case it can be assumed that RQQs show flux variability without varying in polarization. It is clear that we could explain pure intranight variability without polarization variability by viscous instabilities in the accretion disk. We would not expect rapidly changing or high polarization from the accretion disk. Thus these kind of mechanisms are a promising approach to explain intranight variability without accompanying polarization and PINOV. On the other hand, if we could affirm the connection between PINOV and INOV this would clearly point to jet phenomena causing the variability in radio-loud objects.

In any case, there should be intermediate sources, thus studying PINOV and INOV in different classes of AGN should give a lead to the emission mechanism behind intranight variability. However, extremely large samples are needed to overcome scatter in the properties of different samples. 

\section{CONCLUSIONS}

We conducted one of the first systematic studies of polarization intranight variability. In this paper we present observations of a sample of 28 quasars, 12 RQQs and 16 radio-loud blazars, at low to moderate redshifts ($0.048\leq z \leq 1.036$) with a duration of $4$ h each. We used statistical methods to detect polarization intranight variability (PINOV) and measured the degree of polarization.

We find that radio-loud and radio-quiet AGN show clearly different behaviour. None of the RQQs show high levels of polarization. Only two out of 12 RQQs are polarized, both on levels $P < 1$ per cent. On the other hand, 13 out of 16 radio-loud AGN show moderately to highly polarized emission. The three radio-loud AGN that are unpolarized are FSRQs, meaning that all BL Lac type AGN are polarized. Nine out of 15 radio-loud AGN show PINOV. For the FSRQs, four objects show PINOV, whereas one is polarized but non-variable. For the BL Lac type objects five show PINOV, three are polarized but non-variable. We find a potential dependency between degree of polarization and occurency of PINOV with all objects with $P > 5$ per cent showing PINOV, but not all variable objects showing degrees of polarization $> 5$ per cent. We do not find redshift dependencies of the measured degree of polarization, the occurrence of unpolarized emission or the occurrence of PINOV. However, our sample is relatively small, making the results tentative.

We detect two unpolarized objects showing polarization variability, one of them is a RQQ, the other is a FSRQ. Such a behaviour might either be explained by a flickering low level polarization or might be \textquoteleft false\textquoteright polarization variability. If there is underlying constant but low polarized emission from e.g. scattering in the host galaxy or the dust torus, non-polarized varying flux from another component (e.g. the accretion disk) can cause variability in $P_{x/y}$.

We study the violence of the variations in polarization. We do not find the violence of the variations to depend on the degree of polarization. We conclude that there might be a dividing line for variability at $P > 5$ per cent, with all objects showing high degrees of polarization showing PINOV. However our sample is to small to detect a possible dependency between the violence and the degree of polarization. Another possible cause for the seemingly dividing line for PINOV might be the fact that detecting variability in low polarized objects is much more challenging. If we consider a highly polarized object with $P = 25$ per cent a comparatively small relative violence $RV$ is easily detectable, whereas the same relative violence will not be detectable with the same error range at $P \approx 1$ per cent. We detect a probable dependency between the violence and the redshift. Still, this finding is not very definite due to small number statistics. A possible redshift dependency might be either a \textquoteleft real\textquoteright redshift dependency or e.g. wavelength dependency or might even be caused by selection effects as we observe intrinsically brighter objects at higher redshift. 

From our results we conclude that PINOV and highly polarized emission originates from the jet. We conclude that the optical jet emission is either weak or emanating from different, less turbulent regions of the jet in RQQs. We also conclude that whatever causes polarized emission in RQQs is not able to produce high levels of polarization or rapid changes in polarization. The fact that PINOV is relatively common in blazars, especially when high degrees of polarization are observed, poses the question which physical conditions cause this phenomenon. In any case, mechanisms causing PINOV must have very high duty cycles in blazars.

Future carefull study of intranight variability in BL Lacs and FSRQs will have to answer the question what causes the variability. This will be a big step towards understanding the physical conditions in the jet. However, single filter observations might not be enough to reveal the physics behind PINOV. While variations on time scales of weeks and month are already quite well understood and described in some cases (see e.g. \citealt{Ma08}, \citealt{Va08}), short-time polarization scales have been less studied so far. Another point that could not be addressed in this paper is the correlation between INOV and PINOV. However, we hope that this study emphasises the importance of understanding PINOV as it is very common in blazars.

\section*{ACKNOWLEDGEMENTS}

We would like to thank the anonymous referee for helpful comments and constructive criticism. We would also like to thank Tapio Pursimo for numerous suggestions and discussion. These data are based on observations made with the Nordic Optical Telescope, operated on the island of La Palma jointly by Denmark, Finland, Iceland, Norway, and Sweden, in the Spanish Observatorio del Roque de los Muchachos of the Instituto de Astroﬁsica de Canarias. The data presented here have been taken using ALFOSC, which is owned by the Instituto de Astroﬁsica de Andalucia (IAA) and operated at the Nordic Optical Telescope under agreement between IAA and the NBIfAFG of the Astronomical Observatory of Copenhagen. Roy \O{}stensen is supported by the Research Council of the University of Leuven and by the FP6 Coordination Action HELAS of the EU. Jochen Heidt and Janine Pforr acknowledge support by the German Science Foundation (DFG) through SFB439 \textquoteleft Galaxies in the Young Universe\textquoteright.

\label{lastpage}

\end{document}